%% file: morphstore2020.tex
\documentclass{vldb}
\usepackage{graphicx}
\usepackage{paralist}
\usepackage{url}
\usepackage{todonotes}
\usepackage{balance}  % for  \balance command ON LAST PAGE  (only there!)

\vldbTitle{MorphStore: Analytical Query Engine with a Holistic Compression-Enabled Processing Model}
\vldbAuthors{Patrick Damme, Annett Ungeth\"um, Johannes Pietrzyk, Alexander Krause, Dirk~Habich, Wolfgang Lehner}

\vldbDOI{https://doi.org/10.14778/xxxxxxx.xxxxxxx}
\vldbVolume{13}
\vldbNumber{xxx}
\vldbYear{2020}

\begin{document}

% ****************** TITLE ****************************************

\title{MorphStore: Analytical Query Engine with a Holistic Compression-Enabled Processing Model}

% possible, but not really needed or used for PVLDB:
%\subtitle{[Extended Abstract]
%\titlenote{A full version of this paper is available as\textit{Author's Guide to Preparing ACM SIG Proceedings Using \LaTeX$2_\epsilon$\ and BibTeX} at \texttt{www.acm.org/eaddress.htm}}}

% ****************** AUTHORS **************************************


\numberofauthors{1} %  in this sample file, there are a *total*

\author{
% 1st. author
\alignauthor
Patrick Damme, Annett Ungeth\"um, Johannes Pietrzyk, Alexander Krause, Dirk~Habich, Wolfgang Lehner\\
       \affaddr{Database Systems Group}\\
       \affaddr{Technische Universit{\"a}t Dresden}\\
       \affaddr{Dresden, Germany}\\
       \email{\{firstname.lastname\}@tu-dresden.de}
       }
\date{30 July 1999}
% Just remember to make sure that the TOTAL number of authors
% is the number that will appear on the first page PLUS the
% number that will appear in the \additionalauthors section.

\maketitle

\input{00-abstract}

\input{01-introduction.tex}
\input{02-preliminaries}
\input{03-Model}
\input{04-Implementation}
\input{05-evaluation}
\input{07-relatedWork}

\input{08-conclusion}

\section*{Acknowledgment}

This work was partly funded by the German Research Foundation (DFG) within the CRC 912 (HAEC), RTG 1907 (RoSI) as well as by an individual project LE-1416/26-1.

\balance
\bibliographystyle{abbrv}
\bibliography{morphstore2020}  % vldb_sample.bib is the name of the Bibliography in this case

\end{document}

%% file: 00-abstract.tex
\begin{abstract}
In this paper, we present \emph{MorphStore}, an open-source in-memory columnar analytical query engine with a novel \emph{holistic compression-enabled} processing model.
Basically, compression using lightweight integer compression algorithms already plays an important role in existing in-memory column-store data\-base systems, but mainly for base data. 
In particular, during query processing, these systems only keep the data compressed until an operator cannot process the compressed data directly, whereupon the data is decompressed, but not recompressed. 
Thus, the full potential of compression during query processing is not exploited. 
To overcome that, we developed a novel compression-enabled processing model as presented in this paper. 
As we are going to show, the continuous usage of compression for all base data and all intermediates is very beneficial to reduce the overall memory footprint as well as to improve the query performance.
\end{abstract}

%% file: 01-introduction.tex
\section{Introduction}
\label{sec:Motivation}

With increasingly large amounts of data being collected in numerous application areas ranging from science to industry, the importance of online analytical processing (OLAP) workloads increases constantly~\cite{DBLP:journals/cacm/ChaudhuriDN11}. 
OLAP queries typically access a small number of columns, but a high number of rows and are, thus, most efficiently processed by column-stores~\cite{DBLP:journals/cacm/ChaudhuriDN11}.
Moreover, the significant developments in the main memory domain in recent years have rendered it possible to keep even large data sets entirely in main memory.
For these reasons, in-memory column-store database management systems (DBMS) have established themselves as state-of-the-art for OLAP workloads~\cite{DBLP:journals/cacm/BonczKM08,DBLP:journals/ftdb/FaerberKLLNP17,DBLP:conf/vldb/StonebrakerABCCFLLMOORTZ05}.

In these systems, lightweight integer compression algorithms already play an important role~\cite{DBLP:journals/ftdb/AbadiBHIM13,DBLP:conf/sigmod/AbadiMF06,DBLP:journals/ftdb/FaerberKLLNP17,DBLP:conf/sigmod/LangMFB0K16}. 
On the one hand, with the help of some additional lightweight computations for integer compression, the necessary memory space can be reduced~\cite{DBLP:journals/ftdb/AbadiBHIM13,DBLP:conf/sigmod/AbadiMF06,DBLP:journals/ftdb/FaerberKLLNP17,DBLP:conf/sigmod/LangMFB0K16}.
As we have shown in~\cite{DBLP:conf/edbt/DammeHHL17,DBLP:journals/tods/DammeUHHL19}, there is a large variety of lightweight integer compression schemes available and there is no single-best algorithm, but the decision depends on several factors, most importantly on the data characteristic. 
On the other hand, compressed data also offers several advantages for data processing such as less time spent on load and store instructions and a better utilization of the cache hierarchy.
Moreover, a direct processing of compressed integer data is possible in many cases~\cite{DBLP:conf/icde/FengL15,DBLP:conf/sigmod/FengLKX15,DBLP:conf/sigmod/LangMFB0K16,DBLP:journals/pvldb/LeeABCDHI0LLMMPQRSSSZ14,DBLP:conf/sigmod/LiP13,DBLP:journals/pvldb/WillhalmPBPZS09}.
For example, several column scan approaches have been presented in the literature, where filter predicates are directly evaluated on compressed integer data~\cite{DBLP:conf/sigmod/FengLKX15,DBLP:conf/sigmod/LiP13,DBLP:journals/pvldb/WillhalmPBPZS09}.
However, existing systems only provide a very limited set of compression algorithms for base data~\cite{DBLP:journals/ftdb/AbadiBHIM13,DBLP:conf/sigmod/AbadiMF06,DBLP:journals/ftdb/FaerberKLLNP17,DBLP:conf/sigmod/LangMFB0K16}.
Furthermore, during query processing, these systems only keep the data compressed until an operator cannot process the compressed data directly, whereupon the data is decompressed, but not recompressed~\cite{DBLP:journals/ftdb/AbadiBHIM13,DBLP:conf/sigmod/AbadiMF06,DBLP:journals/ftdb/FaerberKLLNP17,DBLP:conf/sigmod/LangMFB0K16}. 
Thus, the full potential of compression during query processing is not exploited from our point of view.

\begin{figure}
    \centering
    \includegraphics[width=\linewidth]{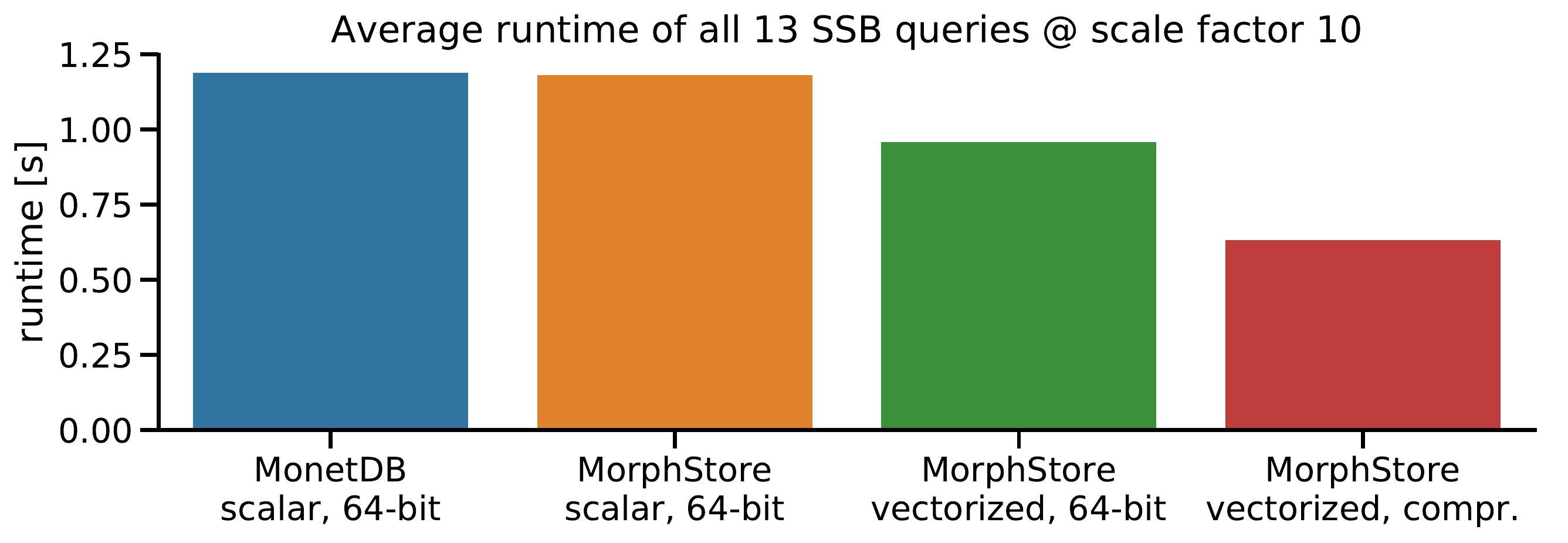}
\caption{Comparing MonetDB and \emph{MorphStore} using SSB~\protect\cite{DBLP:journals/corr/Sanchez16a}. More details in Section~\ref{sec:Evaluation}.}
    \label{fig:teaser}
    \vspace{-0.7cm}
\end{figure}

\textbf{Core Contribution.}
To exploit this potential as shown in Figure~\ref{fig:teaser}, we designed a novel \emph{holistic compression-enabled processing model} satisfying four design principles.  
\begin{compactenum}
\item[\textbf{DP1}] Our model is a compression-aware optimization of the well-known operator-at-a-time processing model introduced by MonetDB~\cite{DBLP:journals/cacm/BonczKM08,DBLP:journals/debu/IdreosGNMMK12}.
Thus, all intermediate results should be representable using a lightweight integer compression algorithm. 
With that, we want to enable the \emph{continuous} usage of compression for the whole query execution.
\item[\textbf{DP2}] Since data characteristics have an impact on the compression scheme decision and usually change during query processing, a suitable scheme should be chosen for each intermediate from a rich and easily extensible set of schemes. 
Moreover, the selection for each intermediate should not depend on the scheme used for another one to independently adapt to its particular data characteristics. 
This implies that a change of the compression scheme from one intermediate to the next should be possible in a very efficient and flexible way.
\item[\textbf{DP3}] No physical columnar query operator should require the uncompressed materialization of its entire input or output data, since this would severely limit the benefits achievable through compression. In particular, a full decompression of the input data should be avoided.
\item[\textbf{DP4}] Related work in this domain is manifold.
Thus, our novel approach should build on this and should extend this work so that the design principles DP1 to DP3 can be realized. 
That also means that existing work should be seamlessly integrable leading to a holistic approach. 
\end{compactenum}

To prove the benefits of our \emph{holistic compression-enabled} processing model, we developed \emph{MorphStore}, an open-source analytical query engine written in C++~\cite{DBLP:conf/sigmod/HabichDUPKHL19}.
As highlighted in Figure~\ref{fig:teaser}, \emph{MorphStore} provides an average runtime behavior comparable to MonetDB~\cite{DBLP:journals/cacm/BonczKM08,DBLP:journals/debu/IdreosGNMMK12} for all SSB queries in case of single-threaded scalar processing of uncompressed $64$-bit data. 
Generally, to improve the query performance, vectorization using SIMD (Single Instruction Multiple Data) instructions is state-of-the-art in this domain~\cite{DBLP:conf/sigmod/FengLKX15,DBLP:journals/pvldb/KerstenLKNPB18,DBLP:conf/sigmod/PolychroniouRR15,DBLP:conf/damon/PolychroniouR19,tvl,DBLP:conf/sigmod/ZhouR02}.
Thus, with a vectorized processing of uncompressed data in \emph{MorphStore} using Intel's latest SIMD extension AVX-512, we are able to decrease the average query runtime by $19$\%. 
However, with our \emph{compression-enabled model} executed with AVX-512 as well, we achieve an average query runtime reduction of about $54$\%, while reducing the memory footprint by $52$\% at the same time. 

\textbf{Contributions in Detail and Outline.}
Thus, to present \emph{MorphStore} with the novel \emph{holistic compression-enabled} processing model in detail, we make the following contributions:
\begin{compactenum}
\item We systematically summarize preliminary work in Section~\ref{sec:Preliminaries}.
These works are manifold but there is no approach to holistically combine their potentials. 
\item Based on that, we describe our novel processing model in Section~\ref{sec:Model} by presenting four different degrees of integration opportunities of compression and operators.
\item In Section~\ref{sec:Engine}, we introduce our analytical query engine \emph{MorphStore} as an implementation for our novel processing model. In particular, we discuss some implementation aspects in more detail. 
\item Then, we present selected results of our exhaustive evaluation of \emph{MorphStore} using micro-benchmarks and the SSB~\cite{DBLP:journals/corr/Sanchez16a} in Section~\ref{sec:Evaluation}. As we are going show, the continuous usage of compression for query execution is very beneficial from two perspectives: reducing \emph{memory footprint} and reducing \emph{query runtime}.  
\item Our evaluation shows that the choice of the compression scheme for intermediates is a new dimension for query optimization. 
Thus, we highlight some initial steps for a compression-aware optimization in Section~\ref{sec:Evaluation}. 
\end{compactenum}

Finally, we present related work in Section~\ref{sec:RelatedWork} and briefly summarize the paper in Section~\ref{sec:Conclusion}.

%% file: 02-preliminaries.tex
\section{Preliminaries}
\label{sec:Preliminaries}

Without claim of completeness, we start with a review of related work on light\-weight integer compression for in-memory column-store systems on: (i) lightweight compression algorithms, (ii) query operators for compressed data, and (iii) available integration approaches of compression into query execution. 
Based on that, we clearly motivate our novel \emph{holistic compression-enabled} processing model.

\subsection{Lightweight Integer Compression}
\label{sec:Narrow:Algos}

Generally, each base data attribute is stored separately as a sequence of values~\cite{DBLP:journals/ftdb/AbadiBHIM13,DBLP:conf/sigmod/AbadiMF06,DBLP:journals/ftdb/FaerberKLLNP17}.
To reduce the necessary memory space, the columns are compressed with a common approach~\cite{DBLP:journals/ftdb/AbadiBHIM13,DBLP:conf/sigmod/AbadiMF06,DBLP:journals/ftdb/FaerberKLLNP17}: (i) encode the values of each column as a sequence of integers using some kind of dictionary encoding~\cite{DBLP:conf/sigmod/BinnigHF09} and (ii) apply lightweight lossless integer compression to each sequence of integers resulting in a sequence of compressed column codes~\cite{DBLP:journals/ftdb/AbadiBHIM13,DBLP:conf/sigmod/AbadiMF06,DBLP:conf/edbt/DammeHHL17,DBLP:journals/ftdb/FaerberKLLNP17}. 
For the second point, a large corpus of lightweight integer compression schemes has been developed and we have to distinguish between techniques, algorithms, and implementations~\cite{DBLP:conf/edbt/DammeHHL17,DBLP:journals/tods/DammeUHHL19,DBLP:journals/spe/LemireB15}.

\textbf{Techniques:} Five basic techniques are currently known and frequently used: frame-of-reference (FOR)~\cite{DBLP:conf/icde/GoldsteinRS98,DBLP:conf/icde/ZukowskiHNB06}, delta coding (DELTA)~\cite{DBLP:journals/spe/LemireB15,DBLP:journals/sigmod/RothH93}, dictionary encoding (DICT)~\cite{DBLP:conf/sigmod/AbadiMF06,DBLP:conf/sigmod/BinnigHF09,DBLP:journals/sigmod/RothH93,DBLP:conf/icde/ZukowskiHNB06}, run-length encoding (RLE)~\cite{DBLP:conf/sigmod/AbadiMF06,DBLP:journals/sigmod/RothH93}, and null suppression (NS)~\cite{DBLP:conf/sigmod/AbadiMF06,DBLP:journals/sigmod/RothH93}.  
FOR and DELTA represent each value as the difference to a certain given reference value or to its predecessor value, respectively. 
DICT replaces each value by its unique key given by a dictionary.
The objective of these three techniques is to represent the original data as a sequence of small integers, which is then suited for actual compression using the NS technique. 
NS is the most well-studied technique and its basic idea is the omission of leading zeros in the bit representation of small integers. 
Finally, RLE tackles uninterrupted sequences of occurrences of the same value, so-called runs. 
In its compressed format, each run is represented by its value and run length. 
Thus, the compressed data is a sequence of such pairs. 
To sum up, FOR, DELTA, DICT, and RLE address the logical level, while NS considers the physical level. 

\textbf{Algorithms:} The genericity of these basic techniques is the foundation to tailor lightweight integer compression algorithms to different data characteristics~\cite{DBLP:conf/edbt/DammeHHL17,DBLP:journals/tods/DammeUHHL19}.
Thus, a concrete algorithm can be described as a cascade of one or more of these basic techniques.
On the algorithm level, NS has been studied most extensively \cite{DBLP:journals/ir/AnhM05,DBLP:journals/spe/AnhM10,DBLP:journals/ws/DelbruCT12,DBLP:journals/tit/Elias75,DBLP:journals/spe/LemireB15,DBLP:journals/ipl/LemireKR18,DBLP:journals/corr/PlaisanceKL15,1090789_Rice,DBLP:conf/damon/SchlegelGL10,DBLP:conf/cikm/SilvestriV10,DBLP:conf/cikm/StepanovGREO11,DBLP:conf/www/YanDS09,DBLP:conf/www/ZhangLS08,DBLP:journals/tois/ZhaoZLSNYW15,DBLP:conf/icde/ZukowskiHNB06}.
An example NS algorithm is \emph{Binary Packing (BP)}~\cite{DBLP:journals/spe/LemireB15,DBLP:conf/cikm/SilvestriV10}. 
The basic idea of BP is to partition a sequence of integer values into blocks and compress every value in a block using a fixed bit width (namely, the effective bit width of the largest value in the block). 
The logical-level techniques, however, have not received much attention \emph{on the algorithm level}.
Different algorithms for RLE differ, e.g., in the way they record the repetitions.
The run value can be stored together with the run length \cite{DBLP:journals/isci/LemireK11}, or the run's start position in the input sequence \cite{DBLP:journals/isci/LemireK11}, or both \cite{DBLP:conf/sigmod/AbadiMF06}.
Furthermore, runs of length one, which allow no actual reduction of the number of data elements, could be stored in a special way to avoid the overhead of storing the run length \cite{DBLP:journals/isci/LemireK11}.
The preprocessing steps DELTA, FOR, and DICT have usually been investigated in connection with the NS technique \cite{DBLP:journals/ws/DelbruCT12,DBLP:conf/icde/ZukowskiHNB06}.

\textbf{Implementation:}
An \emph{implementation} is a hardware-specific executable code of an algorithm.
In recent years, the efficient \emph{vectorized} implementation using SIMD instruction set extensions (Single Instruction Multiple Data) to reduce the runtime has attracted most of the attention~\cite{DBLP:journals/spe/LemireB15,DBLP:journals/corr/PlaisanceKL15,DBLP:conf/damon/SchlegelGL10,DBLP:conf/cikm/StepanovGREO11,DBLP:conf/icde/UngethumPDHL18,DBLP:journals/tois/ZhaoZLSNYW15}. 
The focus of those vectorized implementations is first and foremost on NS algorithms.
For example, SIMD-BP~\cite{DBLP:journals/spe/LemireB15} is a vectorized implementation of the NS algorithm BP.
In comparison, the logical-level techniques have been neglected, although there are also some papers presenting vectorized implementation for DELTA~\cite{DBLP:journals/spe/LemireB15} or RLE~\cite{DBLP:journals/tods/DammeUHHL19,DBLP:conf/icde/UngethumPDHL18}.  

\textbf{Experimental Survey.}
In~\cite{DBLP:conf/edbt/DammeHHL17,DBLP:journals/tods/DammeUHHL19}, we presented an exhaustive experimental evaluation of this large corpus of light\-weight integer compression algorithms and implementations. 
Our analysis using the evaluation metrics compression rate, compression speed, decompression speed, and aggregation speed led us to the overall conclusion that there is no single-best lightweight compression algorithm, but the choice is non-trivial and depends on the following factors: \mbox{(i) data} characteristics, (ii) hardware properties, (iii) SIMD extension and (iv) objective. 
Moreover, we found out that the best algorithm with respect to the compression rate is not necessarily optimal regarding (de)compression speed. 
This allows interesting trade-offs between these two fundamental objectives for optimization.

\subsection{Query Operators for Compressed Data}
\label{sec:Narrow:Operators}
A major advantage of lightweight integer compression is that some query operators can process the compressed data directly, without decompression. 
For that, there is a lot of related work available and we can classify them into processing data compressed at the logical and the physical level.

\textbf{Logical Level:}
DICT was the first compression scheme whose qualification for direct processing was investigated.
In~\cite{143840_graefe}, the authors identified that the equality-preserving property of DICT can be exploited by query operators employing exact-match comparisons. 
For instance, selection scans can process dictionary keys directly, without looking up the actual values in the dictionary, by mapping the selection predicate’s constants to keys using the same dictionary as for the data.
Furthermore, they note that duplicate elimination, grouping, equality joins, and set operations can directly work on dictionary keys. 
Especially for equality joins and set operations, the authors explicitly assume the same dictionary across different columns of the same domain.

In contrast to that, Lee et al.~\cite{DBLP:journals/pvldb/LeeABCDHI0LLMMPQRSSSZ14} proposed to encode each attribute individually to exploit skew in the data distribution for an efficient encoding. 
However, this implies that join-operators can no longer compare dictionary keys from different inputs directly any more.
To address this issue for hash joins, they propose to perform a so-called \emph{encoding translation} by re-encoding the keys of the build-side with the dictionary of the probe-side. 
After the dictionary encodings have been reconciled, dictionary keys, i.e., compressed values are inserted into the hash table to avoid decompression in both the build phase and the probe phase.

Additionally, Abadi et al.~\cite{DBLP:conf/sigmod/AbadiMF06} sketched how database operators can directly process run-length encoded data.
In detail, they mention that an RLE-compressed inner of a nested loop join does not need to be decompressed.
Instead, the data elements in the outer need to be compared only to the run values of the inner.
Each time a comparison succeeds, multiple join matches are found at once, whereby the number of matches is the corresponding run length in the inner.
Furthermore, a summation on a run-length encoded column can be done by summing up the products of corresponding pairs of run value and run length.

\textbf{Physical Level:}
Most DBMSs support different data types for integral attributes and offer physical query operators tailored to these types.
For instance, the SQL standard defines the types \texttt{TINYINT}, \texttt{SMALLINT}, \texttt{INTEGER}, and \texttt{BIGINT}, which represent a single value using 1, 2, 4, and 8 bytes, respectively. 
Representing data in one of these types is, perhaps, the simplest form of lightweight compression.  
Thus, it can be seen as a particular NS format, namely a form of BP, whereby a common bit width is used for all elements of a column and this bit width must be either 8, 16, 32, or 64, making the compressed data byte-aligned.
Thus, most DBMSs can implicitly process data compressed directly.

Indeed, even recent research works on compressed processing frequently restrict the physical-level compression to byte-aligned NS~\cite{DBLP:conf/sigmod/AbadiMF06,DBLP:journals/pvldb/BarberLPRSACLS14, DBLP:conf/sigmod/LangMFB0K16,DBLP:journals/sigmod/WestmannKHM00}. 
This has two reasons: (i) byte-alignment suits the byte-addressability of main memory naturally and (ii) data elements of 8, 16, 32, and 64 bits can be processed
natively on current microprocessors using both classical scalar and modern vectorized
instructions.
The second reason is especially important, since it implies that all operators can be designed to directly work on compressed data with a low implementation effort.
However, some authors observed that such simple byte-aligned packing approaches lack support for arbitrary bit widths and work at a sub-optimal bit-level parallelism~\cite{DBLP:conf/sigmod/FengLKX15,DBLP:conf/sigmod/LiP13,DBLP:journals/pvldb/WillhalmPBPZS09}. 
Thus, some query operators have been proposed for more sophisticated physical-level compression formats.

In this direction, the column scan has received a lot of attention. 
For example, Willhalm et al. proposed \emph{SIMD-Scan}, a full column scan algorithm for data packed with arbitrary bit widths~\cite{DBLP:journals/pvldb/WillhalmPBPZS09}.
They observed that such full column scans are usually memory-bound, but become compute-bound when performed on compressed data.
To alleviate this, they focused on a vectorized implementation using Intel’s SSE instructions on 128-bit vector registers. 
Li and Patel criticize in~\cite{DBLP:conf/sigmod/LiP13} that the \emph{SIMD-Scan} algorithm suffers from a sub-optimal bit-level parallelism, since the employed 32-bit comparisons effectively waste available bits, if the bit width is below 32. 
To address this issue, they propose to use so-called \emph{bit-parallel methods}. 
In particular, the authors introduce two pairs of a physical memory layout and a column scan algorithm efficiently processing data in this layout using an algebraic framework. 
Feng et al.~\cite{DBLP:conf/sigmod/FengLKX15} build upon the work of Li and Patel and especially focus on using SIMD instructions.
They call their approach \emph{ByteSlice} by proposing a new layout and a suitable scan algorithm.
Besides selection, also aggregation has been studied for data compressed at the physical level. 
For example, Feng and Lo proposed bit-parallel implementations of common aggregation
functions directly on compressed data~\cite{DBLP:conf/icde/FengL15}. 

\subsection{Compression and Query Execution}

Besides individual operators for compressed data, there are different ways to employ compression in a query as a whole.
Chen et al.~\cite{DBLP:conf/sigmod/ChenGK01} intensively investigated the integration of compression into query execution in the context of disk-centric row-stores and introduced three strategies:
\begin{compactitem}
\item[\textbf{Eager decompression:}] When base data is loaded into main memory, it is immediately decompressed and the entire query processing takes place on uncompressed data.
\item[\textbf{Lazy decompression:}] Here, the data is kept in compressed form during query processing as long as possible.
That is, base data and, perhaps, early intermediate results are represented in a compressed way and processed by compression-enabled operators.
However, as soon as an operator cannot process the compressed data directly, the data is decompressed and from this point, all processing happens entirely on uncompressed data, which incurs unnecessarily large intermediates and wastes the potential of working on compressed data directly.
\item[\textbf{Transient decompression:}] Here, operators incapable of processing compressed data directly decompress their inputs \emph{only temporarily} and \emph{keep the compressed input elements}.
The processing is done on uncompressed data elements, but for the output, the compressed input elements are used again, such that subsequent query operators can still benefit from compression.
\end{compactitem}

For in-memory column-store systems, the \emph{lazy decompression} strategy has been investigated in several works~\cite{DBLP:journals/pvldb/DasYZVVKGKM15,DBLP:conf/sigmod/LangMFB0K16,DBLP:journals/pvldb/RamanABCKKLLLLMMPSSSSZ13,DBLP:conf/icde/ZukowskiHNB06}. 
For example,~\cite{DBLP:conf/sigmod/LangMFB0K16} proposes a storage format for cold (analyical) data subdividing a column into blocks, each of which can be represented in its individual compressed format.
The formats to choose from are a variant of RLE as well as FOR and DICT combined with a byte-aligned NS algorithm.
Based on that, they present an approach to integrate a vectorized column scan into the compilation-based query engine of HyPer~\cite{DBLP:conf/icde/KemperN11}. 
In particular, their scan outputs a list of logical positions in the compressed column, which are subsequently extracted and decompressed.
After that, the query execution continues with uncompressed data.
Another example is Oracle's in-memory engine \cite{DBLP:journals/pvldb/DasYZVVKGKM15} executing only selection scans on compressed data as well. 

A more sophisticated approach was presented in~\cite{DBLP:conf/sigmod/AbadiMF06}.
The authors show how they integrate five compressed formats into the query execution of the column-store \emph{C-Store} \cite{DBLP:conf/vldb/StonebrakerABCCFLLMOORTZ05}.
They observe that supporting $n$ compressed formats requires $n$ variants of all unary query operators and $n^2$ variants of all binary query operators.
Since this amounts to a high total number, their main focus is on the reduction of the integration effort.
Most importantly, each base and intermediate column is split into blocks that can be accessed by operators via a special API.
This API abstracts from the particular format by exposing properties exploitable by query operators, e.g., whether the data in the block is sorted or whether the block contains only one distinct value.
Consequently, query operators are not specialized to individual formats, but to such properties, which can reduce the number of variants.
As a fallback, the API allows to decompress the block so that the operator can iterate over the uncompressed data elements.
The authors' extension to C-Store also supports compressed intermediates.
However, an operator's output format is hard-coded for each (combination of) input format(s) and chosen depending on what seemed easy to implement to the authors.
In particular, data characteristics of the intermediates are not taken into account.

In Hyrise \cite{DBLP:conf/vldb/Boissier18,DBLP:conf/edbt/0001J19,DBLP:conf/edbt/DreselerK0KUP19}, base data is represented as columns horizontally partitioned into segments, whereby each segment can have its individual compressed format.
To limit the effort of integrating compression into the query execution, the authors also decide to introduce a layer of abstraction. 
In particular, they implement iterator-based methods for sequential and random access to each compressed format.
Query operators make use of the general iterator-interface irrespective of the underlying compressed formats of their inputs.
Intermediate results are not explicitly compressed.

The only work we are aware of that explicitly investigates the compression of intermediate results is \cite{DBLP:journals/vldb/GuzunC16}.
Unfortunately, the authors focus only on complex queries over \emph{bit vectors}, i.e., another data structure, with the operators \texttt{AND}, \texttt{OR}, and \texttt{XOR}.
Furthermore, they distinguish only between uncompressed and compressed data, but not between different compressed formats.
Nevertheless, their motivation is similar to ours:
They also observe that the characteristics of intermediate bit vectors may change during query processing, rendering either the compressed or the uncompressed representation more suitable on a \emph{per-intermediate} basis.
Therefore, they want to support operators on all possible combinations of (un)compressed inputs and outputs.
While they can reuse (un)compressed-only operators from previous works, they contribute variants for mixed compressed and uncompressed inputs.
Moreover, they decouple the output format from the input formats by reusing existing methods to append to (un)compressed bit vectors.

\subsection{Lessons Learned}

Lightweight integer compression is widely used by state-of-the-art systems \cite{DBLP:conf/sigmod/AbadiMF06,DBLP:journals/pvldb/DasYZVVKGKM15,DBLP:conf/edbt/DreselerK0KUP19,DBLP:conf/sigmod/LangMFB0K16,DBLP:journals/pvldb/RamanABCKKLLLLMMPSSSSZ13}, whereby the degree of integration varies.
The approaches to integrate compression into query operators range from a generic and transparent decompression during the reading data access~\cite{DBLP:conf/icde/ZukowskiHNB06} over abstractions for compressed formats \cite{DBLP:conf/sigmod/AbadiMF06} to highly format-specific algorithms for particular operators \cite{DBLP:conf/icde/FengL15,DBLP:conf/sigmod/FengLKX15,DBLP:conf/sigmod/LiP13,DBLP:journals/pvldb/WillhalmPBPZS09}.
However, the main focus of compression is always on the storage and processing of \emph{base data}.
The lightweight compression of \emph{intermediate results} has only been considered for row stores \cite{DBLP:conf/sigmod/ChenGK01} and bit vectors \cite{DBLP:journals/vldb/GuzunC16}.
To the best of our knowledge, a \emph{systematic} investigation of lightweight integer compression for intermediate results in complex analytical queries in in-memory column-stores has never been addressed before.
Thus, we close this gap with our \emph{holistic compression-enabled} processing model. 

%% file: 03-Model.tex
\section{Compression-enabled Model}
\label{sec:Model}

The overall goal of our \emph{holistic compression-enabled processing model} is to enable the continuous use of lightweight integer compression for intermediate results in in-memory column-stores while pursuing the design principles as introduced in Section~\ref{sec:Motivation}.
In this sense, our novel model can be seen as an optimization of the \emph{operator-at-a-time} processing model pioneered by MonetDB~\cite{DBLP:journals/cacm/BonczKM08,DBLP:journals/debu/IdreosGNMMK12}.
The operator-at-a-time model explicitly materializes intermediates, because each operator within an query execution plan (QEP) is evaluated to completion over its entire input data, before subsequent data-dependent operators are invoked~\cite{DBLP:journals/debu/IdreosGNMMK12}. 
In this section, we present the main concepts of our novel model by describing (i) the foundations, (ii) the compression-enabled operators and (iii) the integration into the query execution. 

\subsection{Foundations}

First of all, base data, intermediate results, and query results of our compression-enabled processing model are of exactly the same nature, whereby the elements of each column are unsigned integers. 
Moreover, all base data and intermediates are fully materialized in compressed form using an arbitrary lightweight integer compression algorithm, whereby each column has exactly one format. 
Alternatively, a column can also be uncompressed if \emph{desired}.
If the integers were obtained through a dictionary coding, we assume an individual dictionary per domain as proposed in~\cite{DBLP:journals/pvldb/LeeABCDHI0LLMMPQRSSSZ14}. 
If range predicates need to be evaluated, we assume the dictionary coding to be order-preserving. 
Otherwise, the equality-preserving property of DICT suffices without any changes to our processing model.

Second, our query operators are strongly inspired by those of MonetDB~\cite{DBLP:journals/vldb/BonczK99}.
Initially, MonetDB’s operators processed BATs consisting of a head and a tail column~\cite{DBLP:journals/cacm/BonczKM08,DBLP:journals/debu/IdreosGNMMK12}, but meanwhile, these operators have been re-engineered to work on headless BATs, i.e., mere sequences of values~\cite{MonetDBGoesHeadless}. 
We adapt this latter approach, since it fits our needs of processing sequences of integers very well. 

\begin{figure}
    \centering
    \includegraphics[width=0.9\linewidth]{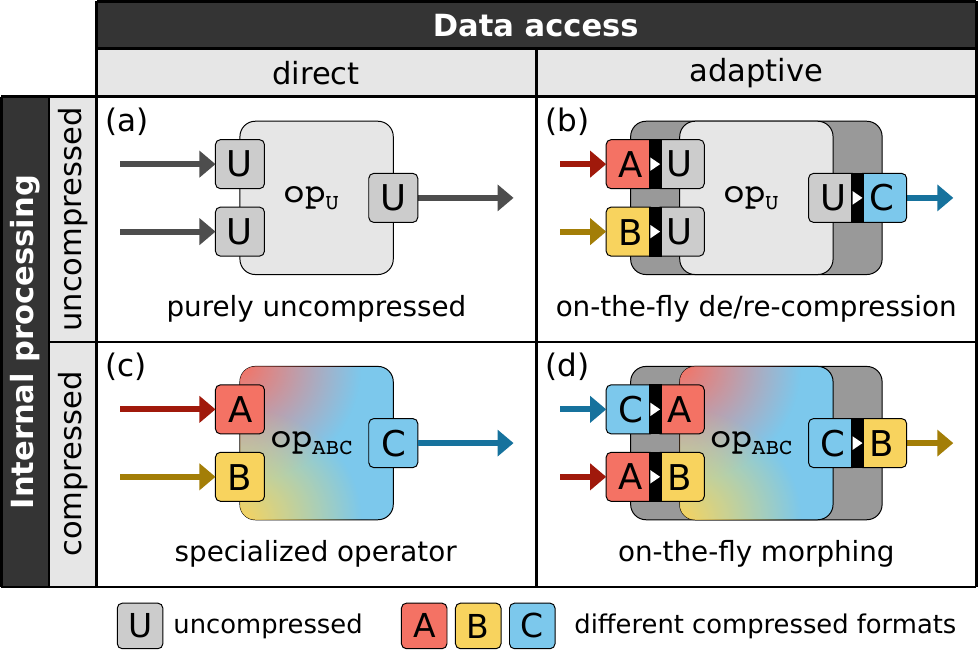}
    \caption{Degrees of integrating lightweight integer compression into query operators at the example of an operator \texttt{op} with two inputs and one output.}
    \label{fig:proc:pm:degrees}
    \vspace{-0.5cm}
\end{figure}

\subsection{Compression-enabled Operators}

A major challenge for our \emph{holistic compression-enabled processing} model is the interplay between operators and (de)compression. 
For that, we distinguish four degrees of integration possibilities and these are derived from two orthogonal dimensions as illustrated in Figure~\ref{fig:proc:pm:degrees}.
The first dimension addresses the \emph{internal processing}:
The operator could either execute its operations on \emph{uncompressed} data elements or on data elements \emph{compressed} according to a particular format, whereby the operator could expect an individual format for each of its input and output columns.
The second dimension concerns the operator's \emph{data access}:
Here, the first option is to access the data \emph{directly} in the format it is represented in.
This is only possible if this format matches the format of the operator's internal processing.
Alternatively, the operator could \emph{adapt} the data to the formats it expects internally.
Input data needs to be adapted before it is consumed and output data is first produced in the operator's internal format and afterwards adapted to the desired output format.
Each combination of these two dimensions is possible and each of them has its individual advantages and disadvantages with respect to both implementation effort and typical optimization objectives such as performance and memory footprint.
In the following, we explain each degree of integration in more detail.

\subsubsection*{Purely Uncompressed}

This trivial degree (Figure~\ref{fig:proc:pm:degrees}(a)) is characterized by an internal processing of \emph{uncompressed} data with \emph{direct} data access.
It involves no compressed data at all and serves merely as a baseline for the following concepts.

\subsubsection*{On-the-Fly De/Re-Compression}

Query operators adopting \emph{on-the-fly de/re-compression} can handle compressed inputs and outputs (adaptive data access), but process uncompressed data internally (Figure \ref{fig:proc:pm:degrees}(b)).
This is currently the most important degree of integration for us, since it can easily be implemented, but already fulfills the design principles DP1 to DP3.
Nevertheless, it leaves the potential of working directly on compressed data unexploited, such that additional degrees are sensible.

This degree is achieved by a wrapper around the operator which temporarily decompresses the inputs and recompresses the outputs of the operator.
This idea is not entirely new and it was inspired by \emph{transient decompression} proposed by Chen et al.~\cite{DBLP:conf/sigmod/ChenGK01}.
However, \emph{on-the-fly de/re-compression} goes one step further than transient decompression, since it allows a recompression to \emph{another format} than the input.
The employed wrapper can work at different levels of granularity at both the input and the output side. 

The coarsest granularity is the \textbf{entire column}.
Using this approach for all operators in a QEP does not make sense, since this would mean that an operator first produces an entire uncompressed column, which is then compressed as a whole, but afterwards decompressed as a whole again by the following operator.
A more reasonable granularity is the \textbf{L$x$-cache-resident block}.
The basic idea is not to materialize uncompressed data in main memory, but only in the cache hierarchy.
On the input side, the wrapper decompresses a block of compressed data when required and hands only the uncompressed block to the operator.
On the output side, the operator first stores uncompressed data elements, which are recompressed when the block size is reached.
Hence, (de)compression and operator execution are separated at the function/routine-level.
On both sides, the block size does not need to match the compressed format's block size (if the format works with blocks at all).
Instead, the block size should be chosen in a cache-conscious way to guarantee that the decompressed data is still in the L$x$-cache when the processing starts and the data to be recompressed is still in the L$x$-cache when the wrapper issues the recompression of a block.
Note that this approach is related to RAM-cache decompression proposed by Zukowski et al.~\cite{DBLP:conf/icde/ZukowskiHNB06} with the difference that our \emph{on-the-fly de/re-compression} is designed to also output compressed data, thereby allowing for compressed intermediates.

The finest granularity is the \textbf{vector register}.
This is the most fine-grained option possible with a state-of-the-art vectorized processing~\cite{DBLP:conf/sigmod/FengLKX15,DBLP:journals/pvldb/KerstenLKNPB18,DBLP:conf/sigmod/PolychroniouRR15,DBLP:conf/damon/PolychroniouR19,tvl,DBLP:conf/sigmod/ZhouR02}.
On the input side, the wrapper decompresses the data elements one vector at a time and forwards single vectors of uncompressed data elements to the operator.
On the output side, the operator produces vectors of uncompressed data elements, which it passes to the wrapper for instantaneous recompression.
This means that the borders between (de)compression and operator execution are blurred, as these are separated only at the instruction-level.
Unfortunately, this granularity is not possible for all combinations of \emph{output} formats and operators.
Some lightweight integer compression algorithms need to analyze a certain number of data elements to decide how to compress them.
For instance, SIMD-BP128 \cite{DBLP:journals/spe/LemireB15} needs to determine the maximum bit width of a block of 128 data elements before it can pack the data.
This bit width cannot be known when just one vector register of uncompressed data elements is available at a time.

\textbf{Advantages:}
This degree already enables a continuous use of compression for all intermediates (and base data) with a low integration effort. 
To support $n$ compressed formats for one operator, the original operator for uncompressed data is reused.
Additionally, only $n$ compression and $n$ decompression algorithms are required, which can be used for wrapping other logical operators as well.
Moreover, the decoupling of the decompression of the inputs and the recompression of the outputs allows the formats of all intermediates to be chosen independently from each other, such that a change of the compressed format during the query execution is easily possible by configuring the wrapper accordingly.

\textbf{Disadvantages:}
The obvious downside is its internal processing of \emph{uncompressed} data.
This does not only limit the possible data-level parallelism, but also prohibits a more sophisticated exploitation of the particular compressed formats which could simplify the operators' work.

\subsubsection*{Specialized Operators}

A \emph{specialized} operator works on compressed data internally (Figure \ref{fig:proc:pm:degrees}(c)).
It is tailored for a specific combination of compressed formats of its inputs and outputs and accesses these directly.
Thus, the intermediates serving as the inputs and outputs must be represented in exactly the formats expected by the operator.
Nevertheless, it can improve the performance by processing compressed data without \emph{any} decompression and through an increased data-level parallelism.
Furthermore, it allows to employ operators on compressed data from the literature as presented in Section~\ref{sec:Narrow:Operators} and thus, fulfills our design principle D4.
For example, the proposed column scan approaches BitWeaving~\cite{DBLP:conf/sigmod/LiP13} or ByteSlice~\cite{DBLP:conf/sigmod/FengLKX15} are specialized operators in this sense. 

\textbf{Advantages:}
Specialized operators avoid the materialization of uncompressed data altogether.
Owing to this, they are a promising approach for improving the operator performance compared to on-the-fly de/re-compression.
Upon the bandwidth saving achieved through the avoidance of uncompressed data transfer between main memory and CPU, specialized operators add a reduction of the computational effort caused by the integration of lightweight integer compression.
This is achieved by the avoidance of any (de)compression overhead, the increase of the data-level parallelism, and the exploitation of format-specific information to shortcut the operator execution.

\textbf{Disadvantages:}
The most decisive drawback of specialized operators is the high conceptual and integration effort they incur.
To support all combinations of $n$ compressed formats for the $i$ inputs and $o$ outputs of one operator, $n^{i+o}$ variants of that operator must be provided.
Unless all of these variants are available, the choice of the intermediates' formats is restricted, since they must match those expected by the operators.
This is a significant burden to fulfilling design principle DP2.
Therefore, we propose to employ specialized operators only selectively.
Finally, a specialization also implies that a particular operator might only be applicable and beneficial in rare cases, depending on the data characteristics of the intermediate and the QEP structure.

\subsubsection*{On-the-Fly Morphing}

Operators employing \emph{on-the-fly morphing} process compressed data internally, but can handle inputs and outputs in \emph{different} compressed formats by means of an adaptive data access (Figure \ref{fig:proc:pm:degrees}(d)).
Similar to \emph{on-the-fly de/re-compression}, this adaptation of the formats is performed by a wrapper.
However, this wrapper does not employ (de)compression algorithms, but so-called \emph{direct morphing algorithms}, which are capable of changing the data representation from one compressed format to another one~\cite{DBLP:conf/adbis/DammeHL15a}.
In fact, the idea of adapting the compressed formats of an operator's inputs has already been sketched by Lee et al.~\cite{DBLP:journals/pvldb/LeeABCDHI0LLMMPQRSSSZ14}.
However, their \emph{encoding translation} addresses only (i) the inputs (ii) of join-operators, and (iii) only DICT-compressed data.
In contrast, our proposed on-the-fly morphing~\cite{DBLP:conf/adbis/DammeHL15a} (i) can also be applied to an operator's outputs, (ii) is possible for any query operator, and (iii) supports arbitrary compressed formats.
Thus, this integration degree enables more flexibility when using specialized operators, since it allows intermediates to have any format, independently from the operator.

Analogous to on-the-fly de/re-compression, the wrapper enabling on-the-fly morphing can work at the same three levels of granularity and we only briefly comment on noteworthy differences compared to on-the-fly de/re-compression.
Morphing at the granularity of the \textbf{entire column} is still not favorable.
However, it does not introduce fresh \emph{uncompressed} intermediates any more, since these are now also \emph{compressed}.
Regarding the granularity of the \textbf{L$x$-cache-resident block}, instead of materializing \emph{uncompressed} data in the caches, with on-the-fly morphing only \emph{compressed} data is materialized.
Nevertheless, the cache-conscious materialization can be expected to be beneficial for compressed data as well.
Morphing is also possible at the \textbf{vector register} granularity \cite{DBLP:conf/adbis/DammeHL15a}.
However, the restrictions with respect to the operator's output remain unchanged.

\textbf{Advantages.}
On-the-fly morphing unifies the virtues of on-the-fly de/re-compression and specialized operators.
It provides the full flexibility regarding the choice of the intermediates' formats and yields the benefits of processing compressed data internally.
Using existing specialized operators as a basis, the integration effort is fair:
To support all combinations of $n$ compressed formats for the inputs and output of one operator, $n^2 - n$ direct morphing algorithms are required, one for each ordered pair of two distinct formats.
These morphing algorithms can be reused for all operators.

\textbf{Disadvantages.}
On-the-fly morphing also comes with a certain computational overhead caused by the wrapper, which might or might not be beneficial, depending on the situation and the optimization objective.
Moreover, the direct morphing algorithms have to be developed. 

\subsection{Integration into Query Execution}

A QEP exploiting compression is constructed using our \emph{compression-enabled} query operators in the same manner as for uncompressed processing since the operators' interfaces have the same semantics and only differ in the compressed formats.
In such a QEP, each base column and intermediate result has a particular compressed format (or could remain uncompressed).
However, it is crucial that the plan is \emph{consistent}.
That means, the format of each input and output of each query operator must match the format of the accessed data.
Furthermore, the final query output column(s) should always be uncompressed, since it cannot be assumed that a client application can interpret compressed data.

It is possible to employ the same degree of integration for all query operators in the QEP.
However, it is perfectly valid to use an individual (combination of) degree(s) for each operator.
The decision of the degree of integration for a particular operator depends on two factors: (i) the availability of the respective compression-enabled variant, and (ii) typical objectives of query optimization, such as memory footprint or query runtime.

%% file: 04-Implementation.tex
\section{MorphStore Engine}
\label{sec:Engine}

The open-source analytical query engine \emph{MorphStore}\footnote{The complete source code is available on GitHub: \\ \url{https://morphstore.github.io}} provides our \emph{holistic compression-enabled} processing model implemented in C++. 
Since vectorization using SIMD instructions is state-of-the-art~\cite{DBLP:conf/sigmod/FengLKX15,DBLP:journals/pvldb/KerstenLKNPB18,DBLP:conf/sigmod/PolychroniouRR15,DBLP:conf/damon/PolychroniouR19,DBLP:conf/sigmod/ZhouR02}, all query operators and (de)compression algorithms in \emph{MorphStore} are also explicitly vectorized, whereby we use SIMD functionality through our recently introduced \emph{Template Vector Library} (TVL) \cite{tvl}.
Our TVL allows to write explicitly vectorized C++ code independently of any particular SIMD extension.
Besides generic data types, e.g., for vector registers, it offers so-called \emph{primitives}.
These are C++ function templates providing typical SIMD functionality, such as loading a vector and numerous operations on vectors.
A smart use of C++ template specialization and inlining allows the compiler to effectively substitute each primitive (i) by the corresponding SIMD intrinsic of a particular SIMD extension or (ii) by a corresponding scalar function \emph{with virtually no runtime overhead} \cite{tvl}.
This allows a single operator implementation, which can be tailored (i) to a chosen SIMD extension or (ii) to a scalar version by simply passing a particular template parameter.
Thus, our following descriptions abstract from any particluar SIMD extension or vector length.

\subsection{Column Storage}

We decided to assume (unsigned) \emph{64-bit} integers as the data type of the uncompressed data, since this is the native word width of most common microprocessors nowadays.
Instead of supporting other typical integer widths, such as 8, 16, and 32 bits, as individual \emph{data types}, we view them as particular \emph{compressed formats}.
Going from 32-bit integers assumed by most of the literature on lightweight integer compression \cite{DBLP:journals/spe/LemireB15,DBLP:conf/damon/SchlegelGL10,DBLP:conf/cikm/StepanovGREO11,DBLP:journals/tois/ZhaoZLSNYW15} to 64-bit integers implies that these existing implementations cannot be reused as-is.
Therefore, we reimplemented some algorithms for 64-bit data elements, while otherwise adhering closely to the algorithms' ideas.
In particular, at the logical level, we currently support (i) DELTA and (ii) FOR, and at the physical level, we currently support the following NS schemes (i) static BP (a variant of BP with one block and fixed bit width for all data elements) and (ii) SIMD-BP \cite{DBLP:journals/spe/LemireB15} (see Section~\ref{sec:Narrow:Algos}).\footnote{The low diversity is only due to the   state of the implementation.}
Additionally, cascades of one logical-level and one physical-level algorithm are possible.

\begin{figure}[t]
\centering
\includegraphics[width=0.95\linewidth]{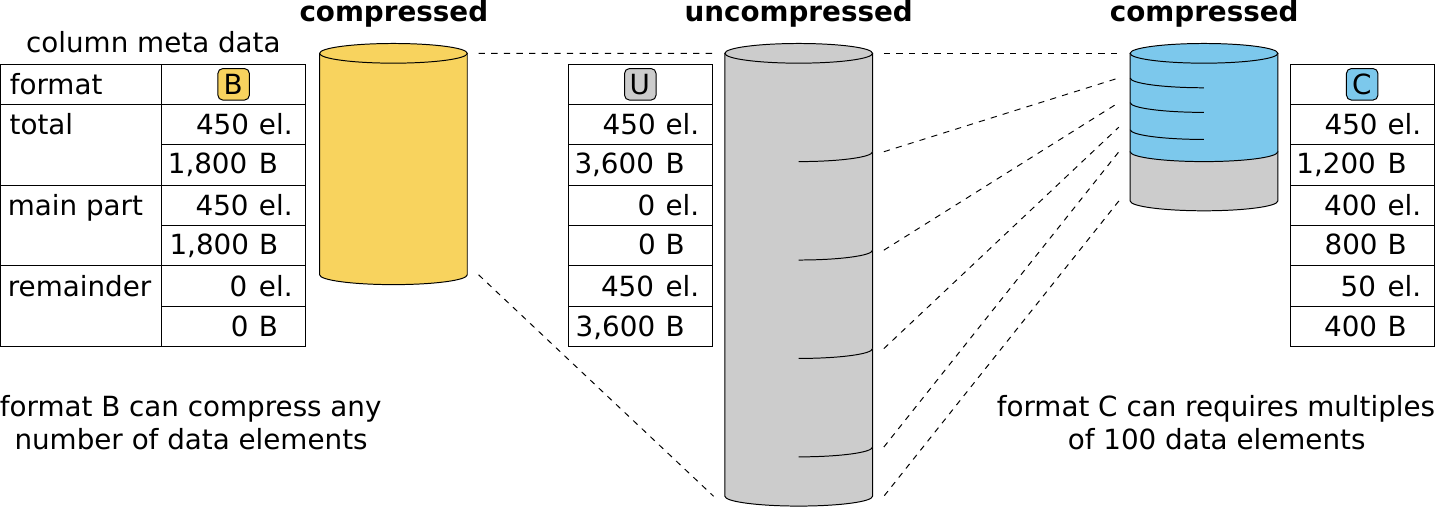}
\caption{
Three representations of the same column.
}
\label{fig:proc:pm:column_layout}
\vspace{-0.4cm}
\end{figure}

Base data, intermediates, and query results are exclusively represented as columns in \emph{MorphStore} and our column data structure is a continuous buffer of bytes. 
Therein, the entire data of the column is stored either uncompressed or compressed according to \emph{exactly one} of the formats mentioned above.
Some lightweight integer compression algorithms are able to compress integer sequences of arbitrary lengths.
However, others subdivide the data into blocks of a certain number of data elements and cannot represent smaller amounts of data, e.g., SIMD-BP512, our port of SIMD-BP128 to AVX-512~\cite{DBLP:journals/tods/DammeUHHL19}, assumes blocks of 512 data elements each.
To be able to deal with columns of arbitrary lengths, each column is subdivided into a compressed main part and an uncompressed remainder as illustrated in Figure~\ref{fig:proc:pm:column_layout}.
Assuming a column of $n$ data elements and a compression algorithm with a block size of $bs$, the compressed part contains the first $\lfloor n / bs \rfloor$ data elements of the column represented in the column's compressed format and the remainder contains the remaining $n \mod bs$ data elements as uncompressed 64-bit integers.
The remainder is stored directly behind the compressed data in the column's buffer and has to be taken into account by operators, too.
A separate structure of meta data stores the sizes of the compressed part and the uncompressed remainder.

\subsection{Query Operators}

Our current implementation is limited to a set of physical query operators that are sufficient to execute the well-established Star Schema Benchmark (SSB)~\cite{DBLP:journals/corr/Sanchez16a}.
Nevertheless, since our presented integration concepts are not operator-specific, other operators could be embraced in the same way. 
Generally, we started with the implementation for a purely uncompressed processing inspired by MonetDB operators~\cite{DBLP:journals/cacm/BonczKM08,DBLP:journals/debu/IdreosGNMMK12}. 
Then, we focused on \emph{on-the-fly de/re-compression} because the continuous use of compression is completely realizable for all analytical queries in a very generic way using that degree of integration. 
Moreover, the \emph{specialized} and \emph{on-the-fly morphing} operators are just further optimizations of \emph{on-the-fly de/re-compression}.  
Generally, a na{\"i}ve implementation of the \emph{on-the-fly de/re-compression} degree would suffer from either (i) low performance due to, e.g., virtual function calls, or (ii) high source code duplication due to the explicit implementation of all combinations, resulting in hardly maintainable source code.
To avoid both of these issues, we apply a number of specific techniques.

As a natural fit to SIMD processing, operating the wrapper for adaptive data access at the vector register-granularity is desirable, since that way, the materialization of uncompressed data even in the caches can be avoided.
Hence, we adopt this approach \emph{on the input side}.
However, due to the reasons mentioned in Section~\ref{sec:Model}, this is not always possible \emph{on the output side}.
Therefore, we use the granularity of an L$x$-cache-resident block here.
Regarding the data access patterns, we can observe that almost all physical operators access both their input and output data in a sequential way.

\begin{figure}[tb]
\centering
\includegraphics[width=\linewidth]{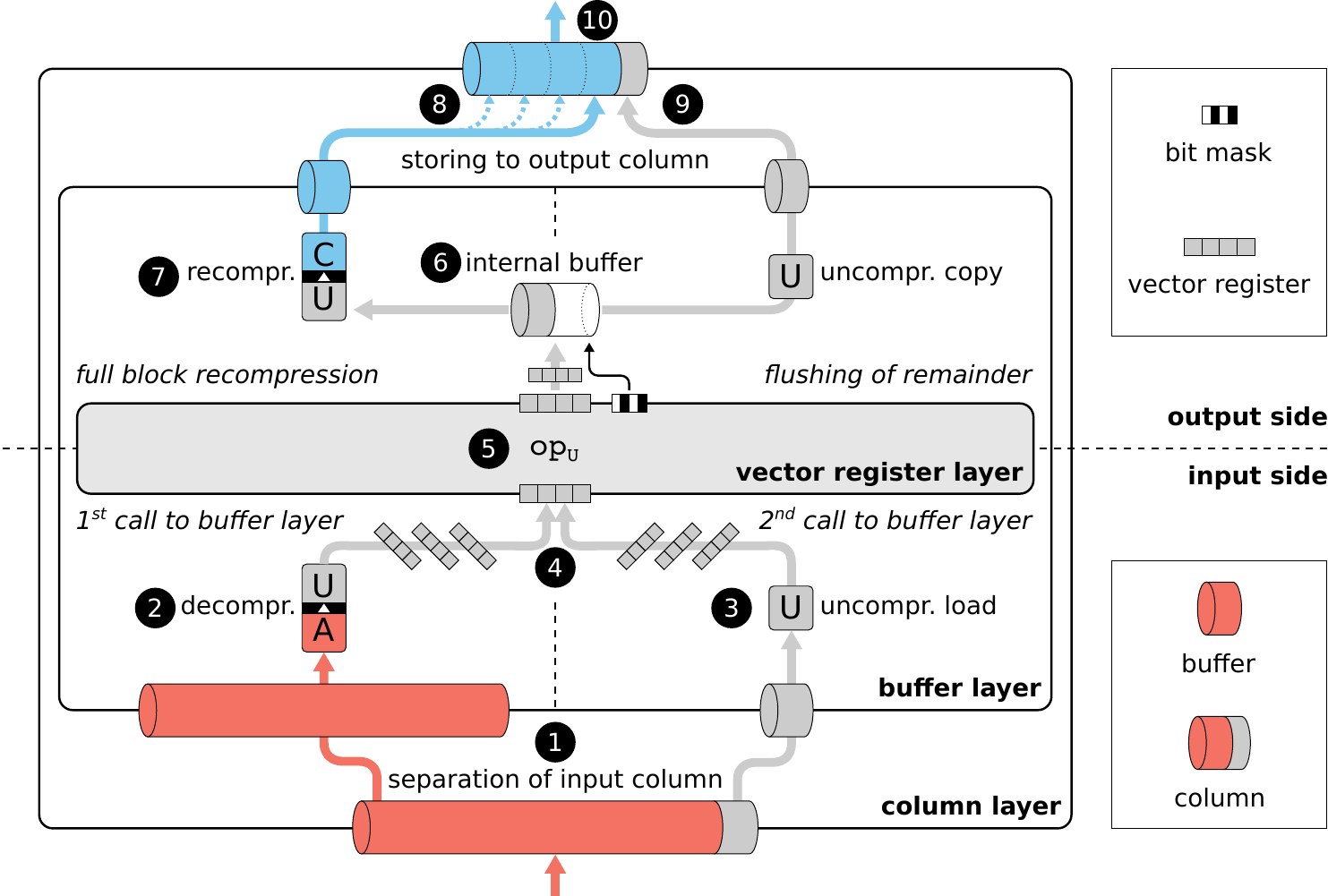}
\caption{
Execution of a query operator adopting on-the-fly de/re-compression.
}
\label{fig:proc:pm:otf_impl}
\vspace{-0.4cm}
\end{figure}

As depicted in Figure \ref{fig:proc:pm:otf_impl}, each \emph{on-the-fly de/re-compression} operator follows a division-of-concerns approach by employing \emph{three layers} with the following responsibilities:
The \emph{column layer} provides the interface to the operator and takes care of the subdivision of the column into a compressed part and an uncompressed remainder.
The \emph{buffer layer} realizes the \emph{wrapper} for the adaptive read and write access to the input and output data, respectively.
Eventually, the \emph{vector register layer} represents the actual \emph{operator core}.
The name of each layer indicates the unit of data it consumes and produces.
Each of these layers is implemented as a C++ function, whereby the column and buffer layers require individual functions for the \emph{input side} and the \emph{output side}.

In the following, we describe the execution of an operator with one input and one output column, both of which are accessed \emph{sequentially}.
With the numbers in brackets, we refer to those in Figure \ref{fig:proc:pm:otf_impl}.
The column layer provides the interface for calling the operator.
On the input side, it identifies the compressed main part and uncompressed remainder of the input column with the help of the column's meta data.
Then, it calls the input-side buffer layer once for each of these two sub-buffers (1).
The input-side buffer layer is essentially the decompression routine of the input column's compressed format (2) or a simple loading of uncompressed data for the remainder (3).
However, instead of storing decompressed vectors to memory, it passes each of them on to the operator core at the vector register layer (4).
The vector register layer consumes vectors of uncompressed 64-bit data elements.
It executes the respective operator on each vector, whereby a vector of uncompressed output data elements is produced (5).
For selective operators, a bit mask indicates which of the output vector's elements are valid.
This output vector, perhaps accompanied by a bit mask, is handed over to the output-side buffer layer.
The output-side buffer layer accepts one uncompressed vector at a time and appends it to its uncompressed internal L$x$-cache-resident buffer, whereby the valid data elements indicated by the bit mask are stored compactly (6).
Once this internal buffer reaches its capacity, the compression algorithm of the output column's format is called (7).
It loads uncompressed data from the internal buffer and appends compressed data to the output column's buffer (8).
After both calls to the input-side buffer layer have terminated, the possible remaining data elements in the output-side buffer layer's internal buffer need to be flushed.
Of this remaining data, as much as possible is appended to the output column in compressed form, while the possible remainder is appended uncompressed (9).
Finally, the output-side column layer returns the output column to the caller of the operator (10).
It is noteworthy that the input-side and output-side work in an \emph{interleaved} fashion, i.e., the output-side is active before the input-side returns.

A close look reveals that the buffer layer is the only layer whose \emph{implementation} depends on the particular format, while the column layer only needs to know the format to invoke the buffer layer correctly, and the vector register layer is not concerned with formats at all.
At the same time, the buffer layer is the only layer that is \emph{not specific to the logical operator}, while the vector register layer is obviously specific, and the column layer needs to know the operator core to pass to the buffer layer.
This constellation enables an economical, non-repetitive implementation based on C++ template metaprogramming.
In particular, the formats of the input and output columns are modeled as template parameters of the column and buffer layers.
The column layer is implemented generically with respect to these formats.
However, for both the input-side and the output-side of the buffer layer, template specializations must be provided for each format to be supported.
These specializations are strongly based on the decompression and compression algorithms of the formats, respectively.
Thus, these specializations can be reused by all operators.
Furthermore, the input-side buffer layer receives the core operator to call as a template parameter as well.
Finally, the column layer has to initialize the vector register layer such that it calls the output format's template specialization of the output-side buffer layer.
The use of templates prevents expensive \emph{virtual} function calls, since the right specializations are determined at compile-time.
Besides that, expensive frequent calls to the vector register layer are avoided by forcing the compiler to inline it into the input-side buffer layer.

However, there are some operators employing random read access or random write access.
For example, the \texttt{project}-operator requires random read access to compressed data, because this operator is used, for instance, to transfer the result of a selection result on one column (sequentical access) to another column (random access). 
Since lightweight integer compression algorithms are designed for efficient sequential access, random access incurs some challenges.
At the logical level of compression, the interpretation of one particular compressed data element might require either meta information, as for FOR and DICT, or even information on all preceding compressed data elements, as for DELTA.
At the physical level of compression, the challenge is threefold: (i) the physical byte or bit address corresponding to a logical position must be determined, (ii) depending on the format, one \emph{or more} random accesses are required to obtain all bits of a code word, and (iii) the original data element must be restored from the obtained bits.
In the literature, random read access has been investigated to \emph{certain compressed formats} \cite{DBLP:conf/sigmod/FengLKX15,DBLP:conf/sigmod/LiP13,DBLP:journals/pvldb/WangLHCPS17}. 
However, we decided to follow a simple approach by restricting random access to BP with a fixed bit width for all data elements (which we call static BP) and uncompressed data.
For these formats, all of the challenges mentioned above can be solved in a straightforward way.
The integration of these formats is again accomplished using one specialization of a template function for random read access per format.
In case of random write access to compressed data, we observed that this very often belongs to the query's result column(s), e.g., in a group-based aggregation. 
Since these shall be uncompressed anyway, we do not yet consider random write access to compressed data.

%% file: 05-evaluation.tex
\section{Experimental Evaluation}
\label{sec:Evaluation}

We conducted our experimental evaluation on a server machine equipped with an Intel Xeon Gold 6130 clocked at 2.1 GHz.
The capacities of the L1, L2, and L3-caches are 32 KiB, 1 MiB, and 22 MiB, respectively.
The system has 4 sockets with 32 cores each and exhibits a non-uniform memory access (NUMA).
However, we only investigate the single-thread performance and ensured that all memory is allocated on the local socket to exclude NUMA effects~\cite{DBLP:conf/btw/KieferSL13,DBLP:conf/sigmod/LeisBK014}.
The size of the ECC DDR4 main memory is 384 GiB and all experiments happened entirely in-memory.
All our operator and (de)compression algorithm implementations are specialized to a scalar and an AVX-512 version through our TVL~\cite{tvl}.
Unless stated otherwise, we report the AVX-512 measurements.
We choose a size of 16 KiB, or 2,048 uncompressed data elements, for the internal buffer used at the output side of our on-the-fly de/re-compression operators.
Note that this is half of the size of the L1 cache of our machine.
We compiled our source code using \texttt{g++} version 8.3.0 with the optimization flag \texttt{-O3}.
The operating system is a 64-bit Ubuntu 18.10 with Linux kernel version 4.18.0-13-generic.
We repeated all time measurements 10 times and report only the means.
Next, we present some micro-benchmarks to provide a clear understanding of \emph{MorphStore}, before we report our results using the Star-Schema Benchmark~\cite{DBLP:journals/corr/Sanchez16a} including a comparison with MonetDB. 

\subsection{Micro-Benchmarks}

\begin{table}
\centering
\caption{
Properties of our synthetic columns.
Each column contains 128 Mi data elements.
}
\label{tab:datach}
\begin{tabular}{|c|l|c|c|}
\hline
 & Data distribution & Sorted & Maximum \\
& & & bit width \\
\hline
\hline
C1 & uniform in $[0, 63]$ & no & 6 \\ % small values
\hline
C2 & 99.99\% uniform in $[0, 63]$ & no & 63 \\ % small values with huge outliers
 & 0.01\% constant $2^{63}-1$ & & \\
\hline
C3 & uniform in $[2^{62}, 2^{62}+63]$ & no & 63 \\ % huge values in narrow range
\hline
C4 & uniform in $[2^{47}, 2^{47} + 100\text{K}]$ & yes & 48 \\ % sorted large values
\hline
\end{tabular}
\vspace{-0.4cm}
\end{table}
The behavior of lightweight integer compression algorithms depends strongly on the data characteristics~ \cite{DBLP:conf/edbt/DammeHHL17,DBLP:journals/tods/DammeUHHL19}.
Since DBMSs have to handle data with arbitrary properties efficiently, we use synthetic columns (C1-C4) for our micro-benchmarks and Table \ref{tab:datach} summarizes their characteristics.

\textbf{Single Operator.}
We first investigate the runtime of a single operator as the building block for complex QEPs.
In particular, we choose the \texttt{select}-operator, which takes a column as input and outputs a sorted column containing the \emph{positions} in the input column that store matching data elements.
Note that these positions are themselves unsigned integers.
For each input column, we want to select the (a-priori known) lowest data element in the column using a point predicate.
We adapt the distributions mentioned in Table \ref{tab:datach}, such that 90\% of all data elements are this value, while the remaining 10\% are distributed as specified.

\begin{figure}
\centering
\includegraphics[width=\linewidth]{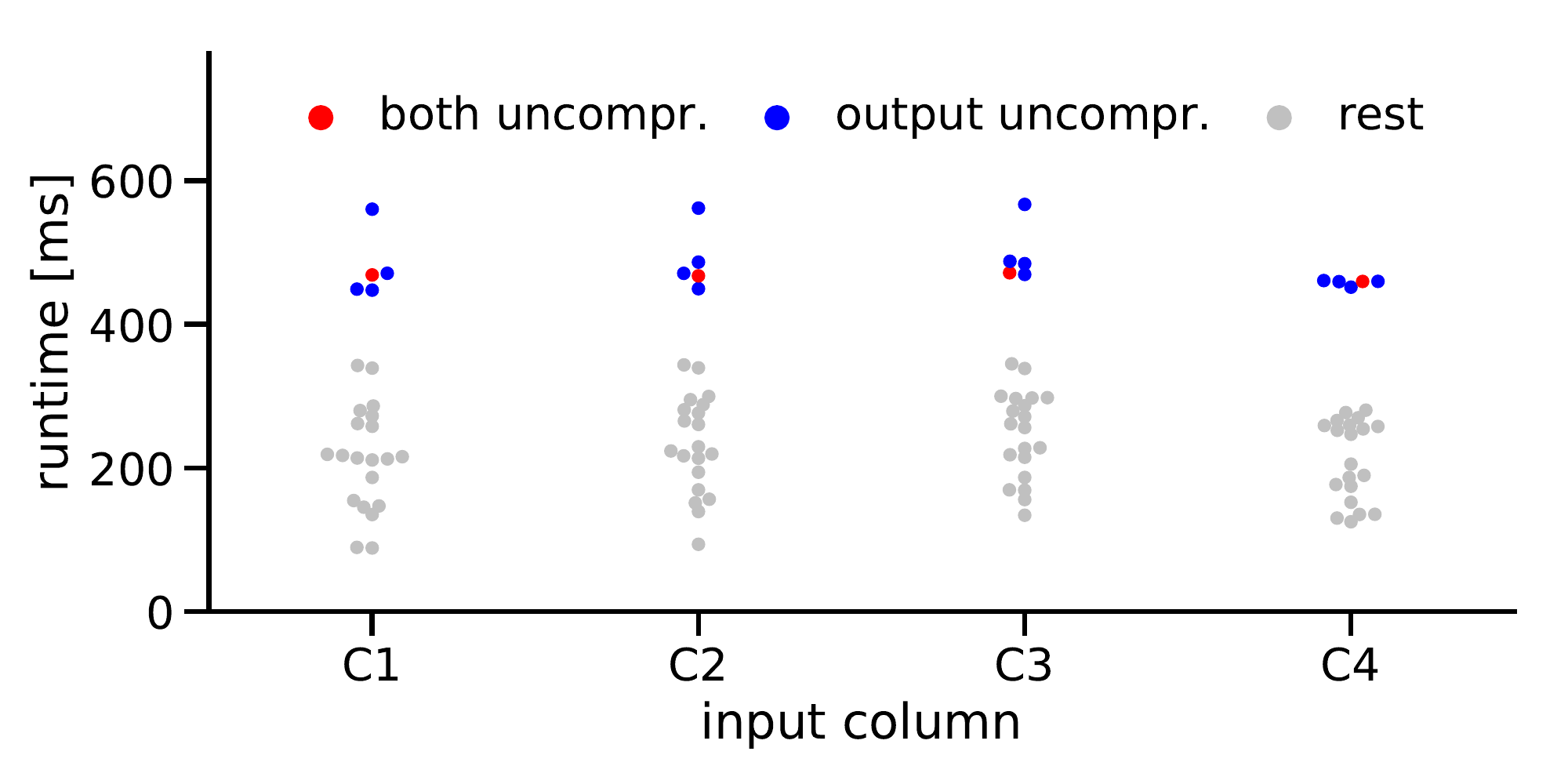}
\caption{
Runtime evaluation \texttt{select}-operator.
}
\label{fig:select}
\vspace{-0.5cm}
\end{figure}

Since \emph{MorphStore} currently supports five compression algorithms, there are 25 format combinations of input and output for the \texttt{select}-operator.
Figure \ref{fig:select} gives an overview of the runtimes of \emph{all} combinations for all input columns (C1-C4).
We see that the runtime of the purely uncompressed processing (red dot) is about the same for all input columns, which is expected.
The employment of our \emph{on-the-fly de/re-compression} (blue and gray dots) can save between 72\% and 81\% of the runtime in the best case, depending on the input column.
At the same time, the runtime can increase by 20\% for C1--C3 in the worst case.
Compressing also the output column (gray dots) can reduce the runtime much further than compressing only the input column (blue dots).
This is interesting, since the output column can only be an intermediate in the context of a QEP.
The input format employed by the fastest combination depends on the input column.
For C1, static BP is preferred, since it contains rather small values.
C2 contains 0.01\% huge outliers, rendering SIMD-BP512 a better choice, since it can adapt to the local data distribution in each block of 512 data values.
The distribution of C3 has a narrow range of huge values, thus, FOR + SIMD-BP512 is most suitable here.
DELTA + SIMD-BP512 results in the best runtime for the sorted column C4.
The output format employed by the fastest variant is DELTA + SIMD-BP512 in all cases, since the output is always sorted.
We conclude that our \emph{on-the-fly de/re-compression} can reduce the runtime of a single operator \emph{significantly}, if the formats are chosen carefully.

\textbf{Simple Query.}
We extend our evaluation to a simple analytical query: given a relation \texttt{R} with attributes \texttt{X} and \texttt{Y}, our query is \texttt{SELECT SUM(Y) FROM R WHERE X = c}.
The first step of the query execution is the selection we have investigated above, i.e., the selectivity is 90\% again, with input column \texttt{X} and output column \texttt{X'}.
After that, a \texttt{project}-operator extracts the data elements at the positions in \texttt{X'} from base column \texttt{Y} producing intermediate \texttt{Y'}.
Finally, a \texttt{sum}-operator aggregates all data elements in \texttt{Y'}.
Thus, our simple query consists of three operators accessing two base columns (\texttt{X} and \texttt{Y}), two intermediates (\texttt{X'} and \texttt{Y'}), and one result column with a single value which we ignore.

\begin{figure}
\centering
\includegraphics[width=\linewidth]{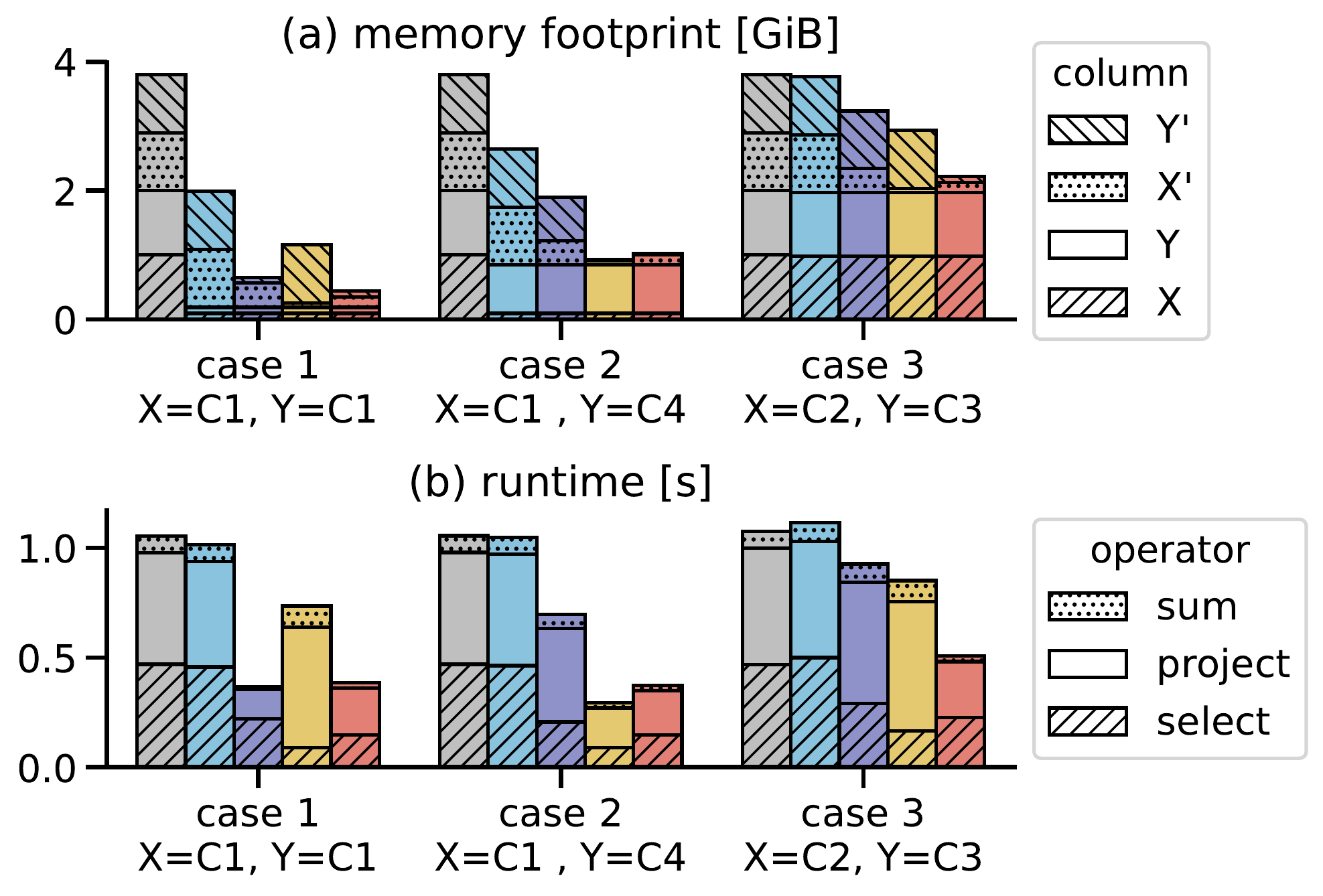}
\includegraphics[width=\linewidth]{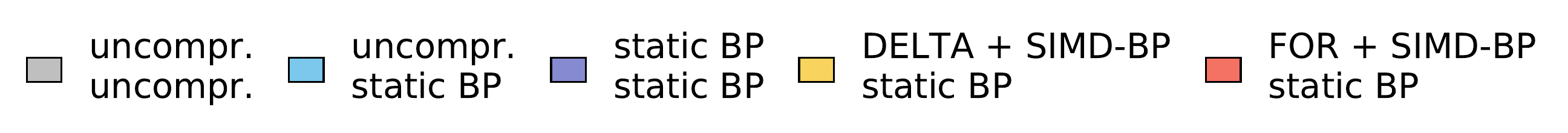}
\caption{Evaluation results for a simple query.}
\label{fig:select_sum}
\vspace{-0.5cm}
\end{figure}

We measured the memory footprint and runtime of this query for three cases, each of which is characterized by a certain combination of the base columns in Table \ref{tab:datach}.
Each column can be assigned an individual compression format and there are $5 \cdot 2 \cdot 5 \cdot 5 = 250$ possible combinations.\footnote{Random access is currently only supported for uncompressed data and static BP.
}
However, here we concentrate on just a few interesting combinations, while not searching for the best one.
Figure \ref{fig:select_sum}(a) shows the results for the query's memory footprint, broken down to the individual columns.
First of all, the footprint of the purely uncompressed processing is the same irrespective of the characteristics of the base columns.
Applying Static BP for the base columns results in a size reduction to 52\% in case 1, since \texttt{X} and \texttt{Y} contain only very small values here.
The other extreme is case 3, where almost no size reduction can be achieved, since both base columns contain data elements of up to 63 bits.
If Static BP is applied to the intermediates as well, further reductions to between 17\% (case 1) and 85\% (case 3) are the consequence.
Representing both intermediates using DELTA + SIMD-BP512 reduces the size of \texttt{X'} significantly in all cases, since this column is always sorted.
At the same time, DELTA + SIMD-BP512 is beneficial for \texttt{Y'} only in case 2, where a reduction to 24\% is achieved, while in case 1, it even yields a worse memory footprint than Static BP resulting in a reduction to only 30\% compared to the uncompressed processing.
In fact, in cases 1 and 3, \texttt{Y'} should rather be represented using \mbox{FOR + SIMD-BP512} to achieve reductions to 11\% and 58\%, respectively.

The runtimes of the whole query and the individual operators are displayed in Figure \ref{fig:select_sum}(b).
Purely uncompressed processing is equally fast in all cases.
Applying Static BP only for the base columns can decrease the runtime by only up to 4\% (case 1), since writing uncompressed intermediates is very expensive.
In case 3, the query runtime is even \emph{increased} by 4\%.
If the intermediates are compressed as well, the runtimes shrink to between 34\% (case 1) and 86\% (case 3).
While using a suitable cascade for the intermediates could reduce the memory footprint in all cases, the runtimes can only be reduced in cases 2 and 3 by using DELTA or FOR cascaded with SIMD-BP512 in both cases.

We conclude that the continuous compression of both base columns and intermediates can reduce the memory footprint as well as the runtime of a query, if the formats are chosen carefully.
Furthermore, given two format combinations, the one that is better with respect to the memory footprint is not always better concerning runtime.

\subsection{Star Schema Benchmark (SSB)}
Next, we investigate the fitness of our novel processing model for complex analytical queries using SSB~\cite{DBLP:journals/corr/Sanchez16a} at scale factor 10. 
We applied an order-preserving dictionary encoding to all string columns in the schema to obtain integer columns.
In fact, all 13 queries can be executed on dictionary keys without looking up the string values.

\begin{figure}
\centering
\includegraphics[width=\linewidth]{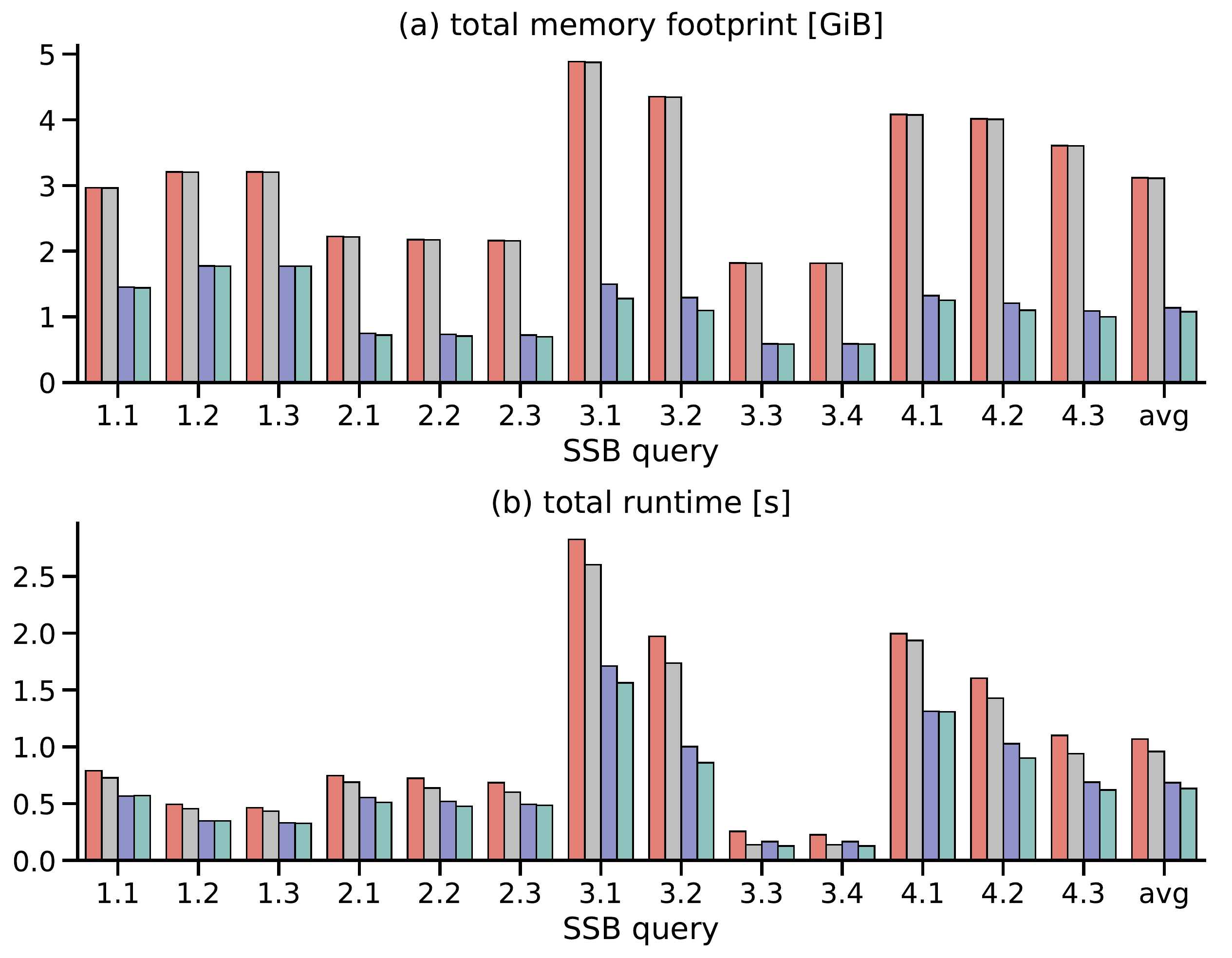}
\includegraphics[width=\linewidth]{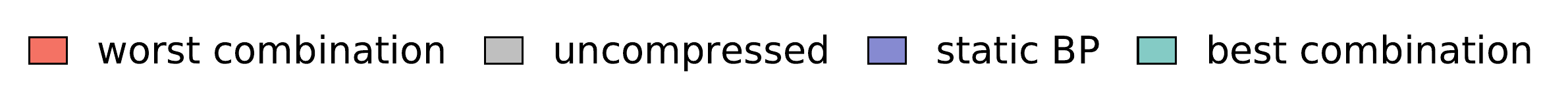}
\caption{
Impact of the format combinations in SSB.
}
\label{fig:ssb_formats}
\vspace{-0.5cm}
\end{figure}

\textbf{Impact of Continuous Compression.} The QEPs of the SSB queries involve between 6 and 16 base columns and between 15 and 56 intermediates.
Each of the columns can be represented in its individual compressed format, which results in a very high number of possible format combinations.
Thus, we first want to find out the impact of different format combinations and the improvement through the optimal format combination.
We allow compression for base columns and intermediates here.
We consider the following format combinations for each query: (i) purely uncompressed, (ii) Static BP for all columns, and (iii/iv) the actual best/worst format combination.
We determined the best/worst combination regarding memory footprint by exhaustively trying each available format for each column \emph{individually} (since column footprints add up) and employing the format yielding the lowest/highest physical size.
Concerning runtime, we applied a greedy strategy which, starting at the base data, considers one column at a time by trying all available formats for that column, measuring the resulting query runtimes and fixing the column's format to the one yielding the best/worst runtime for the next steps of the search.
Note that these best/worst combinations are allowed to employ the uncompressed format.

Figure \ref{fig:ssb_formats}(a) shows the results for the memory footprint.
The purely uncompressed processing achieves by far the worst memory footprints for all queries.
Using Static BP for all columns reduces the memory footprint to between 30\% (Q3.2) and 55\% (Q1.2) (37\% on average).
For Q1.x, 3.3, and 3.4 this is already the best choice.
However, for the remaining queries, the optimal combinations yield a reduction to between 25\% (Q3.2) and 33\% (Q2.2) (35\% on average).
Figure \ref{fig:ssb_formats}(b) depicts the query runtimes.
Here, the worst combination results in a runtime \emph{increase} by 11\% on average, compared to the purely uncompressed case.
Employing Static BP reduces the runtime for all queries except for Q3.3 and 3.4, while the ideal combination reduces the runtimes even for those queries.
In detail, the best runtimes range between 49\% (Q3.2) and 93\% (Q3.3) (66\% on average), compared to purely uncompressed processing.
To sum up, the continuous use of compression can significantly reduce memory footprint and runtime if the formats are well chosen, whereby the potential for reduction is query-dependent.

\begin{figure}
\centering
\includegraphics[width=\linewidth]{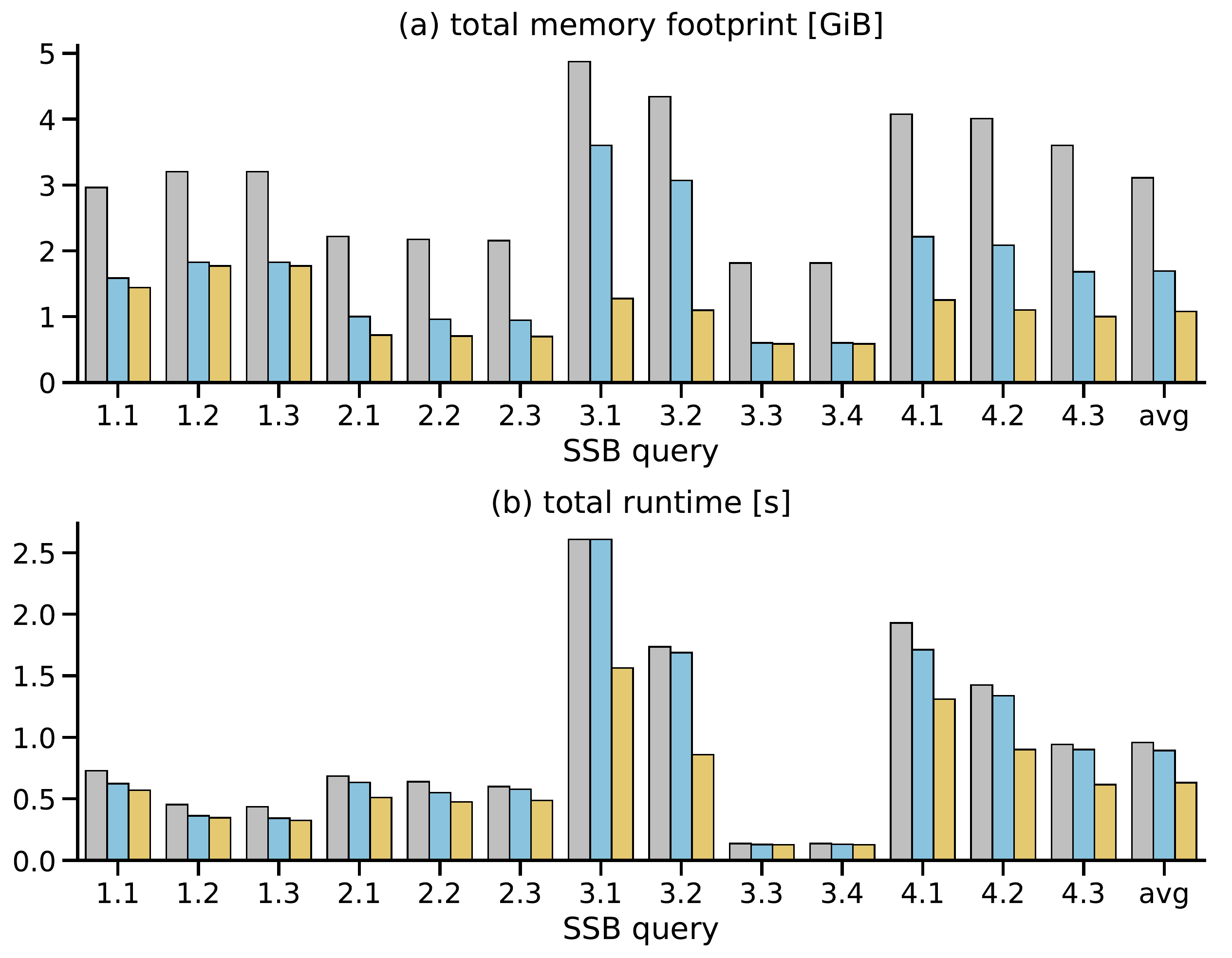}
\includegraphics[width=\linewidth]{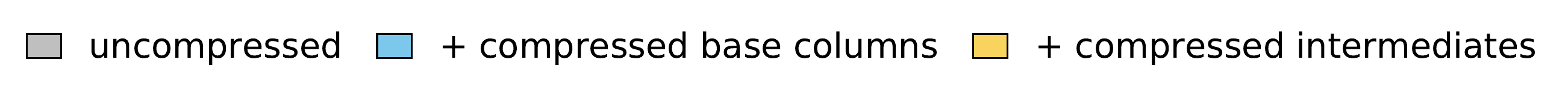}
\caption{Compression base data vs. intermediates.}
\label{fig:ssb_base_vs_interm}
\vspace{-0.6cm}
\end{figure}

\textbf{Impact Compression Base vs. Intermediates.}
Our next question is in how far our continuous compression of \emph{intermediates} contributes to these reductions, compared to the already established compression of \emph{base data}.
Thus, we start by not allowing compression at all, then we allow compression for the base columns only, and, finally, also for the intermediates.
The results for the memory footprint can be found in Figure \ref{fig:ssb_base_vs_interm}(a).
Allowing compression only for the base data already reduces the footprints significantly to between 33\% (Q3.4) and 74\% (Q3.1) (54\% on average).
The impact of offering compression for intermediates too is highly query-dependent.
For Q3.3 and 3.4, no further reduction is possible, while a further reduction to 25\% can be achieved for Q3.2.
On average, compressing intermediates reduces the footprints to 35\%, i.e., 19\% further than with base columns only.
Figure \ref{fig:ssb_base_vs_interm}(b) depicts the runtime results.
These can either be reduced to up to 79\% (Q1.3) or not reduced (Q3.1) when compressing only base data (93\% on average), while the additional compression of intermediates achieves a reduction to between 49\% (Q3.2) and 92\% (Q3.4) (66\% on average).
We can conclude that our continuous compression of intermediates contributes \emph{significantly} to the overall memory and runtime 
reductions achievable through compression.
In fact, regarding the runtime, the intermediates' impact is even higher than that of the base columns, on average.

\textbf{Comparison to MonetDB.} Next, we move on to a comparison to MonetDB, the system that is closest to our processing model \emph{in the uncompressed case}.
To ensure a fair comparison, we compiled MonetDB-11.31.13 with all relevant optimization switches on using the same compiler (and version) and run it on the same machine as \emph{MorphStore}.
We run MonetDB also in single-threaded and read-only mode.
Although MonetDB supports string columns, we use the same dictionary-encoded base data.
For our experiments, the QEPs used in \emph{MorphStore} imitate those of MonetDB as closely as possible, including the same join order.
We used MonetDB's internal tools for measuring the mere query \emph{run}times, excluding the time spent on  query optimization.

\begin{figure}
\centering
\includegraphics[width=\linewidth]{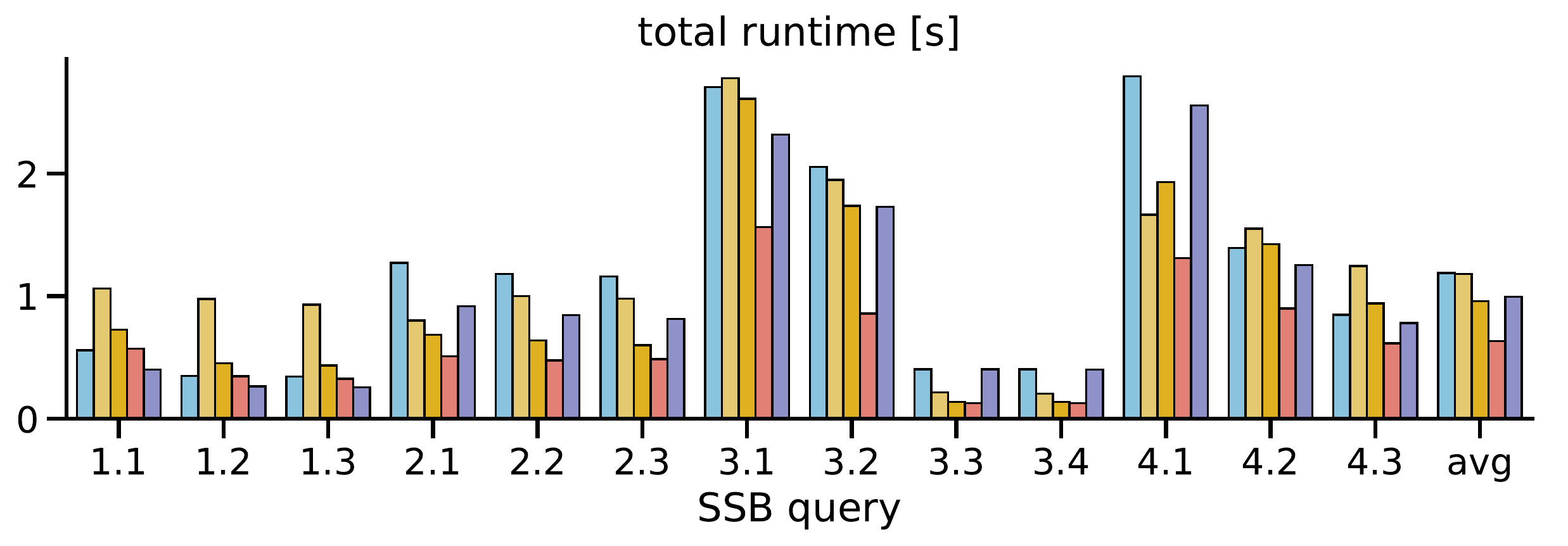}
\includegraphics[width=\linewidth]{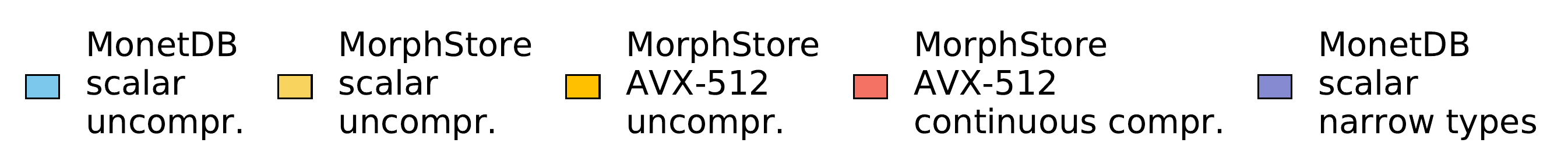}
\caption{
Comparing MonetDB and \emph{MorphStore}.
}
\label{fig:ssb_morphstore_vs_monetdb}
\vspace{-0.5cm}
\end{figure}

The runtime results are given in Figure \ref{fig:ssb_morphstore_vs_monetdb}.
Since MonetDB does not use SIMD instructions, we first compare the scalar execution on purely uncompressed data, i.e., all columns use 64-bit.
None of the systems is faster than the other for all queries and on average, both systems are equally fast.
The vectorized execution using AVX-512 can reduce the runtimes in \emph{MorphStore} by up to 54\% (Q1.2) or increase it by up to 16\% (Q4.1) (on average, it decreases by 19\%).
Also employing our novel continuous compression for base columns and intermediates according to the ideal format combination reduces the runtimes further to between 35\% (Q1.3) and 79\% (Q4.1) of the scalar uncompressed processing in \emph{MorphStore} (54\% on average).
While MonetDB has no explicit support for compression of intermediates, we try to simulate it by using the narrowest integer type possible for all base data columns in MonetDB.
We can see that this improves the runtime of MonetDB to 84\% of its uncompressed runtime, on average, however, this is still much slower than \emph{MorphStore's} \emph{holistic compression-enabled} approach.
Thus, we conclude that the query execution in \emph{MorphStore} achieves runtimes comparable to a state-of-the-art system \emph{in the purely uncompressed case}, while using our novel continuous compression of intermediates in combination with vectorization yields a speed up of $2\times$, on average.

\textbf{Determining a good format combination.}
Above, we have learned that a full exploitation of the potential of our \emph{compression-enabled} processing model requires a good combination of the columns' formats.
Thus, our final question is how to find such a combination efficiently.
In \cite{DBLP:journals/tods/DammeUHHL19}, we have proposed a cost-based selection strategy for lightweight integer compression algorithms, which adopts a gray-box approach by combining explicit modeling of functional properties of the algorithms and calibration capturing hardware-dependent behavior.
Given a particular data set, our cost model can estimate the compression rate as well as (de)com\-pression performance of light\-weight integer compression algorithms and we have proven its effectiveness in \cite{DBLP:journals/tods/DammeUHHL19}.

\begin{figure}
\centering
\includegraphics[width=\linewidth]{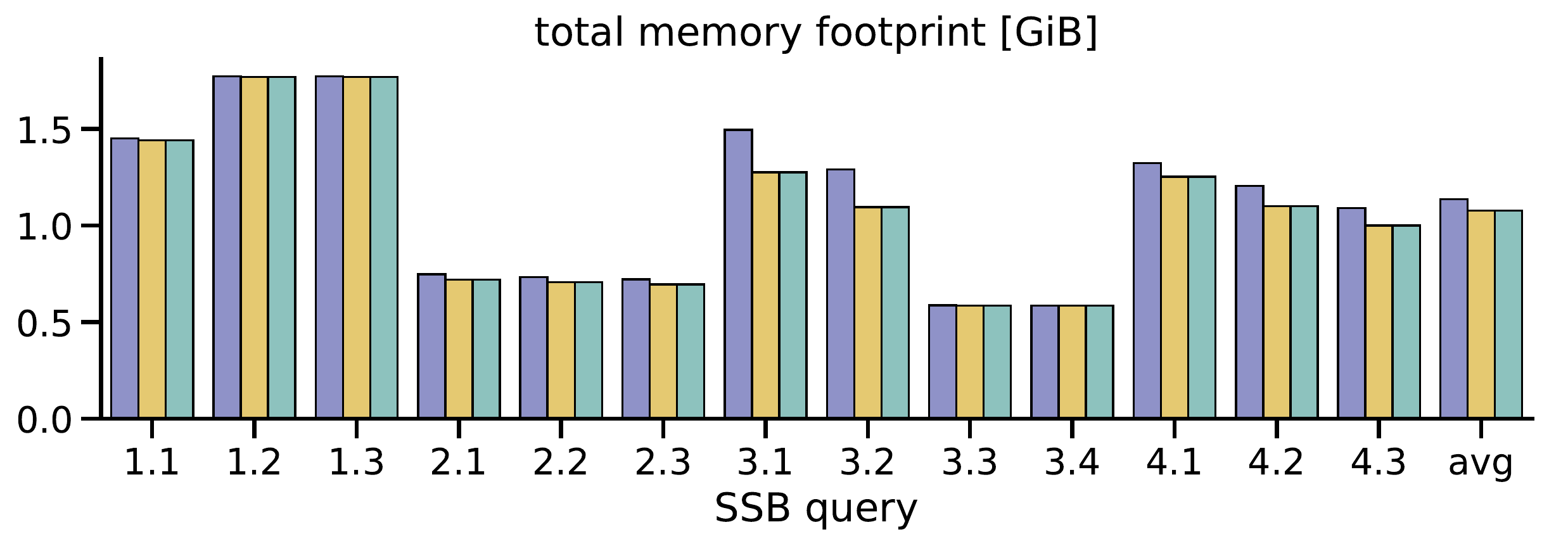}
\includegraphics[width=\linewidth]{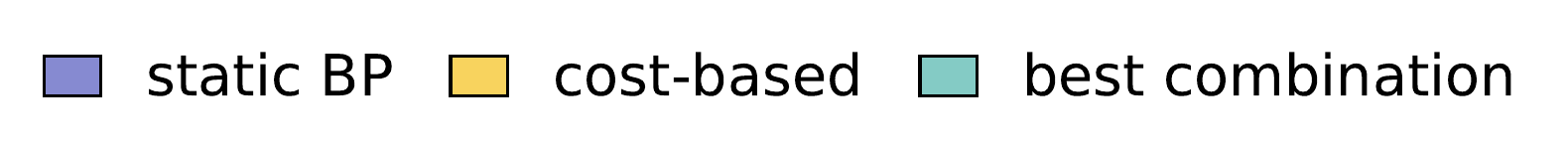}
\caption{
Fitness of our cost-based selection strategy~\protect\cite{DBLP:journals/tods/DammeUHHL19} for obtaining suitable format combinations.
}
\label{fig:ssb_opt}
\vspace{-0.5cm}
\end{figure}

Here, we show that this cost model can be applied to determine a suitable format for each base column and intermediate in a QEP of a complex analytical query.
In particular, we focus on minimizing the memory footprint as one very important optimization objective.
For that purpose, we assume that some basic data characteristics required by our cost model, such as the number of (distinct) data elements, the bit width histogram, and the sort order, are known for all intermediates.
We apply our cost model to obtain a good format combination by selecting a suitable compression algorithm for each base column and intermediate \emph{individually} with respect to the compression rate objective provided by our selection strategy for a single data set as described in \cite{DBLP:journals/tods/DammeUHHL19}.
Figure \ref{fig:ssb_opt} compares the memory footprints resulting from the so-obtained format combinations to those achieved with static BP for all columns as well as the actual best combination found in an exhaustive search.
While na{\"i}vely using static BP for all columns already comes very close to the optimal memory footprints for all SSB queries, the diagram clearly shows that our cost-based selection strategy yields memory footprints virtually equal to the actual optimal ones.
We interpret these results as a promising first step into the direction of complementing our \emph{compression-enabled} processing model with a compression-aware query optimization to select suitable formats for all intermediates.
However, to achieve the ultimate goal of a compression-aware optimization, more research in this direction is essential. 

%% file: 07-relatedWork.tex
\section{Related Work}
\label{sec:RelatedWork}

Throughout the paper, we already covered related work regarding lightweight integer compression, query operators for compressed integer data and integration approaches of compression into query execution.
Now, we focus on the description of different processing models to explain the differences between our novel model and existing ones. 

The classical processing model is the \emph{Volcano}-style iterator-based model (also called \emph{tuple-at-a-time}) proposed by Graefe for disk-oriented row-store database systems~\cite{DBLP:journals/tkde/Graefe94}.
However, this classical model has shown to be unable to make effective use of the facilities of modern processors \cite{DBLP:conf/vldb/AilamakiDHW99,DBLP:conf/iccd/CaoTTLKW99}, especially due to data and instruction cache misses and branch mispredictions.
Further disadvantages are described in~\cite{DBLP:journals/cacm/BonczKM08,DBLP:conf/cidr/BonczZN05,DBLP:conf/cidr/KissingerSHL13,DBLP:journals/pvldb/MenonPM17,DBLP:journals/pvldb/Neumann11,DBLP:conf/icde/PadmanabhanAMJ01,DBLP:conf/sigmod/ZhouR04}.
To overcome these disadvantages and to increase the query performance, a variety of optimizations~\cite{DBLP:conf/icde/PadmanabhanAMJ01,DBLP:conf/icde/PadmanabhanAMJ01,DBLP:journals/tkde/StonebrakerRH90} and completely different processing models~\cite{DBLP:journals/cacm/BonczKM08,DBLP:conf/cidr/BonczZN05,DBLP:journals/tkde/Graefe94,DBLP:conf/cidr/KissingerSHL13,DBLP:journals/pvldb/MenonPM17,DBLP:journals/pvldb/Neumann11,DBLP:conf/icde/PadmanabhanAMJ01,DBLP:conf/sigmod/ZhouR04} have been proposed. 

For example, Kissinger et al.~\cite{DBLP:conf/cidr/KissingerSHL13} suggested an \emph{indexed-table-at-a-time} processing model for in-memory database systems. 
Their initial idea is the full materialization of each intermediate result in the form of an index.
However, since the full materialization is expensive, the authors propose to fuse operators to \emph{composite operators}.
A more sophisticated approach to avoid the materialization of intermediates was introduced by Neumann~\cite{DBLP:journals/pvldb/Neumann11} by maintaining 
the pipeline processing from one operator to the next.
Since these pipelines are query-dependent, they must be compiled at query run-time.
To reduce the implied overhead, the system builds upon the low-level virtual machine (LLVM) compiler framework \cite{DBLP:conf/cgo/LattnerA04} and directly generates the pipeline code in an LLVM intermediate representation.
While the extra effort caused by the query dependent generation is a downside, it also enables the compiler to perform query-specific code optimizations, such that only the instructions required in the particular case need to be executed.
While the overall aim is the avoidance of intermediate results, materialization is still required at the pipeline boundaries, for instance, when building the hash table of a hash join.
The DBMS \emph{Peloton} \cite{DBLP:conf/cidr/PavloAALLMMMPQS17} employs an advancement of this approach that was proposed by Menon et al.~\cite{DBLP:journals/pvldb/MenonPM17}.
The authors argue that the strict avoidance of intermediates and the tuple-at-a-time processing of the approach by Neumann disallow the exploitation of inter-tuple-parallelism offered by modern microprocessors.
In particular, the authors show how to leverage SIMD and prefetching instructions.
Moreover, they perform a strategic partial materialization of selected intermediate results, while otherwise they avoid intermediates using pipelining.

In contrast to the previous approaches, MonetDB~\cite{DBLP:journals/cacm/BonczKM08,DBLP:journals/debu/IdreosGNMMK12} follows a completely different approach. 
Based on the decomposition storage model~\cite{DBLP:conf/sigmod/CopelandK85}, MonetDB uses an \emph{operator-at-a-time} model. 
That means, each operator takes one or more columns as input and produces one or more columns as output.
Hence, all intermediate results are fully materialized. 
Moreover, each operator executes a simple operation on all input elements in a tight loop. 
This solves the problem of frequent instruction cache misses the iterator-based model suffers from, since the bulk of the data is processed by only a few instructions.
Furthermore, it enables compilers to apply optimization techniques such as loop pipelining and makes it easier for the processor to apply out-of-order execution.
Boncz et al.~\cite{DBLP:conf/cidr/BonczZN05,DBLP:journals/debu/ZukowskiBNH05} observe that a downside of the operator-at-a-time is that the full materialization of all intermediates can make the query processing memory-bound if the intermediates' size exceeds that of the cache.
To address this issue, the authors present the \emph{X100} query engine, thereby proposing the \emph{vector-at-a-time} model.
Their goal was to combine the column-wise processing of the operator-at-a-time model with the pipelined execution of the iterator-based model.
The processing is done by so-called \emph{vectorized primitives}.
These are operators which consume and produce partitions of a column, so-called \emph{vectors}.

To sum up, a variety of different processing models have been proposed. 
On the one hand, some of these are designed to avoid the materialization of intermediates whenever it is possible \cite{DBLP:journals/tkde/Graefe94,DBLP:journals/pvldb/Neumann11}.
However, in recent years, systems adopting these processing models have started to relax this aim and perform at least a selective or partial materialization of intermediates to further improve their query performance \cite{DBLP:journals/pvldb/MenonPM17,DBLP:conf/icde/PadmanabhanAMJ01,DBLP:conf/sigmod/ZhouR04}.
On the other hand, some processing models explicitly materialize intermediates either fully \cite{DBLP:journals/cacm/BonczKM08,DBLP:conf/cidr/KissingerSHL13} or partially \cite{DBLP:conf/cidr/BonczZN05}.
In this context, we developed a sensible alternative with our \emph{compression-enabled} processing model that has not yet been considered to this extent.  
In the future, we will enhance our novel model with some pipelining concepts.

%% file: 08-conclusion.tex
\section{Conclusion}
\label{sec:Conclusion}

In this paper, we presented \emph{MorphStore}, an in-memory analytical query engine with a novel \emph{holistic compression-enabled} processing model. 
As we have shown, our novel processing model is an optimization of the well-known \emph{operator-at-a-time} model by establishing a continuous use of integer compression for all intermediates in a query execution plan. 
In particular, we are able to \emph{significantly} reduce the memory footprint and the query runtime for analytical queries.
In addition, our \emph{holistic compression-enabled} model builds on existing work, supplements it with meaningful approaches and combines them in one engine called \emph{MorphStore}.
This holistic approach enables the seamless integration of previous work like BitWeaving~\cite{DBLP:conf/sigmod/LiP13} or ByteSlice~\cite{DBLP:conf/sigmod/FengLKX15} in this domain, but also opens up the possibility of driving the development of specialized operators for processing compressed integer data.
Moreover, new challenges such as determining the best-suited compression scheme for intermediate results will become interesting and this opens ups a completely new dimension for query optimization. 